\documentclass{aa}  

\usepackage{graphicx}
\usepackage{txfonts}
\pdfmapfile{+txfonts.map}
\defcitealias{Shal14}{Paper I}
\defcitealias{Goic16}{Paper II}
\newcommand\avgmstar{\langle M_* \rangle}

\newcommand\avgmhat{\langle M_*/{\rm M_\odot} \rangle}
\begin{document} 

 \title{Resolving the inner accretion flow towards the central supermassive black hole 
 	in SDSS J1339+1310\thanks{Tables 1 and 3$-$4 are only available in electronic 
        form at the CDS via anonymous ftp to cdsarc.u-strasbg.fr (130.79.128.5) or via 
        http://cdsweb.u-strasbg.fr/cgi-bin/qcat?J/A+A/vol/page}}

                                                             
   \author{V. N. Shalyapin\inst{1,2,3}  
     		\and
          	L. J. Goicoechea\inst{1}
		\and
		C. W. Morgan\inst{4}
		\and
		M. A. Cornachione\inst{4}
		\and
		A. V. Sergeyev\inst{3,5}
            }

\institute{Departamento de F\'\i sica Moderna, Universidad de Cantabria, 
    		Avda. de Los Castros s/n, E-39005 Santander, Spain\\
            \email{vshal@ukr.net;goicol@unican.es}
		\and
            O.Ya. Usikov Institute for Radiophysics and Electronics, National 
		Academy of Sciences of Ukraine, 12 Acad. Proscury St., UA-61085 
		Kharkiv, Ukraine
                \and
            Institute of Astronomy of V.N. Karazin Kharkiv National University,
		Svobody Sq. 4, UA-61022 Kharkiv, Ukraine\\
		\email{alexey.v.sergeyev@gmail.com}
		\and
		Department of Physics, United States Naval Academy, 
            572C Holloway Rd., Annapolis, MD 21402, USA\\
		\email{cmorgan@usna.edu;mcornach@gmail.com}
                \and
            Institute of Radio Astronomy of the National Academy of Sciences of 
                Ukraine, 4 Mystetstv St., UA-61002 Kharkiv, Ukraine
		}


 
  \abstract{We studied the accretion disc structure in the doubly imaged lensed quasar 
  SDSS J1339+1310 using $r$-band light curves and UV-visible to near-IR (NIR) spectra from the 
  first 11 observational seasons after its discovery. The 2009$-$2019 light curves
  displayed pronounced microlensing variations on different timescales, and this 
  microlensing signal permitted us to constrain the half-light radius of the 1930 \AA\ 
  continuum-emitting region. Assuming an accretion disc with an axis inclined at 
  $60^\circ$ to the line of sight, we obtained ${\log \left( r_{1/2}/{\rm cm} \right)} 
  = 15.4^{+0.3}_{-0.4}$. We also estimated the central black hole mass from 
  spectroscopic data. The width of the \ion{C}{iv}, Mg\,{\sc ii}, and H$\beta$ 
  emission lines, and the continuum luminosity at 1350, 3000, and 5100 \AA, led to 
  ${\log \left( M_{\rm{BH}}/\rm{M_{\odot}} \right)} = 8.6 \pm 0.4$. Thus, hot gas 
  responsible for the 1930 \AA\ continuum emission is likely orbiting a $4.0\times10^8 
  {\rm M_{\odot}}$ black hole at an $r_{1/2}$ of only a few tens of Schwarzschild 
  radii.}     
  

   \maketitle
%
			
\section{Introduction}
\label{sec:intro}

Microlensing-induced variability in gravitationally lensed quasars allows astronomers  
to determine the sizes of compact continuum-emission regions in distant active galactic 
nuclei \citep[e.g.][]{Wamb98}. Hence, this extrinsic variability has become an 
extremely powerful tool, as evidenced by results for QSO 2237+0305 
\citep[e.g.][]{Koch04,Eige08,Mosq13}. In particular, visible continuum light curves of 
a lensed quasar at redshift $z \sim$ 1$-$2 may provide radii for UV continuum sources 
in the accretion disc around its central supermassive black hole (SMBH). To gain a 
more complete perspective of the inner accretion flow, it is also necessary to measure 
the SMBH mass from spectroscopic data \citep[e.g.][and references therein]{Morg18}. 
For a Schwarzschild black hole, its mass defines the innermost stable circular orbit 
(ISCO) of the gas in the disc, and thus, it provides information about the relative sizes of UV 
continuum sources. Furthermore, using data for 11 lensed quasars, \citet{Morg10} found 
a physically relevant correlation between disc size and SMBH mass. The precision of 
this correlation continues to be improved by the addition of measurements from new lensed 
quasar systems \citep{Morg18,Corn20}. 

\citet{Inad09} reported the discovery of a set of doubly imaged gravitationally lensed 
quasars, which included \object{SDSS J1339+1310}. The two images (A and B) of this 
quasar are located at the same redshift, $z_{\rm{s}}$ = 2.231, while the early-type 
galaxy G at $z_{\rm{l}}$ = 0.607 acts as a gravitational lens \citep{Shal14,Goic16}.
It is also thought that stellar mass objects in G act as gravitational microlenses, 
strongly affecting image B \citep[][henceforth Paper I]{Shal14}. Thus, \object{SDSS 
J1339+1310} is very well suited to the study of its central engine from visible 
continuum light curves and spectroscopic observations. In order to reproduce observed 
microlensing variations via numerical simulations, and measure the SMBH mass, reliable 
extinction and (macro)lens models are also required. These models should rely on 
robust observational constraints, some of which were presented in Table 1 of 
\citetalias{Shal14}. In addition to the astro-photometric solution in the last column 
of that table, we simultaneously obtained the macrolens magnification\footnote{The macrolens 
magnification is usually denoted by $M$, but we use $\mu$ instead to avoid any 
confusion with masses} ratio $\mu_{\rm{B}}/\mu_{\rm{A}}$ = 0.175 $\pm$ 0.015 and the 
dust extinction ratio $\epsilon_{\rm{G,B}}/\epsilon_{\rm{G,A}}$ = 1.33 $\pm$ 0.11 at 
5500 \AA\ in the G rest-frame using some emission line cores and narrow components of carbon 
emission lines that presumably come from very extended regions not affected by microlensing
\citep[linear extinction;][henceforth Paper II]{Goic16}. In \citetalias{Goic16}, we also 
measured the time delay $\Delta t_{\rm{AB}}$ = 
47$^{+5}_{-6}$ d (A is leading), which is a critical quantity for accurately isolating 
the microlensing-induced variability from the source quasar's intrinsic variability.

\begin{figure*}
\centering
\includegraphics[width=15cm]{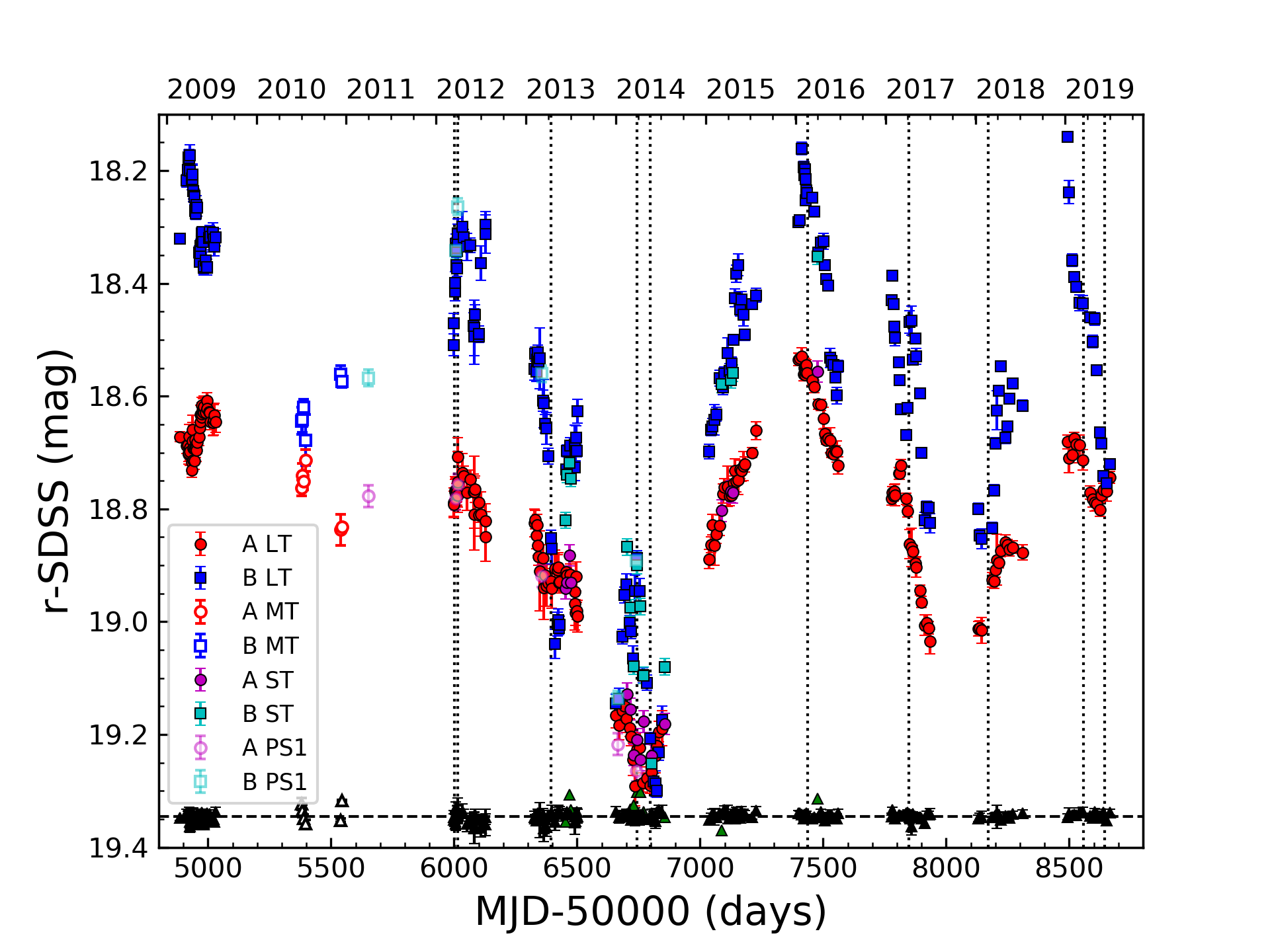}
\caption{Light curves of SDSS J1339+1310AB from its discovery to 2019. These $r$-band 
brightness records of the two quasar images are based on observations from four 
different facilities that are labelled as LT, MT, ST, and PS1 (see main text). 
Magnitudes of a field star are offset by +2.0 (triangles) to facilitate comparison 
with strong quasar variations. Dates of spectroscopic observations are marked by 
vertical dotted lines.}
\label{fig:lcur}
\end{figure*}

In this paper, we mainly focus on the structure of the inner accretion flow in 
\object{SDSS J1339+1310}. Sect.~\ref{sec:obs} presents $r$-band light curves and 
UV-visible to near-IR (NIR) spectra spanning 11 years (2009$-$2019). Our extinction 
and lens models for the system are described in Sect.~\ref{sec:extlens}. 
Sect.~\ref{sec:bhmass} is devoted to determining the mass of the central SMBH, and 
the $r$--band microlensing variability and the size of the corresponding continuum 
source are discussed in Sect.~\ref{sec:microsize}. Our conclusions appear in 
Sect.~\ref{sec:final}. Throughout the paper, we use a flat cosmology with $H_0$ = 70 
km s$^{-1}$ Mpc$^{-1}$, $\Omega_M$ = 0.3, and $\Omega_{\Lambda}$ = 0.7 
\citep{Hins09}.

\section{Observational data}
\label{sec:obs}

\subsection{Light curves in the r band}
\label{sec:lcur}

Within the framework of the Gravitational LENses and DArk MAtter (GLENDAMA) 
project\footnote{\url{https://gravlens.unican.es}}, whose objective is to perform and 
analyse observations 
of a sample of ten gravitationally lensed quasars \citep{Gilm18}, we are conducting an 
$r$-band monitoring of \object{SDSS J1339+1310} with the 2.0 m Liverpool Telescope 
\citep[LT;][]{Stee04}. Light curves basically covering the first visibility season 
after the discovery of the double quasar, as well as four additional seasons in 
2012$-$2015, were presented in \citetalias{Goic16}. Here, we add four new seasons of 
LT data (2016$-$2019) and combine our LT light curves with magnitudes from other 
facilities. It is noteworthy that these complementary data are used to fill the 
2010$-$2011 gap in LT brightness records and improve the sampling in other monitoring 
years. 

The updated light curves of \object{SDSS J1339+1310} consist of 241 observing epochs 
(nights). On the LT, we used the RATCam Charge-Coupled Device (CCD) camera and the Sloan 
$r$ filter with a
central wavelength (CWL) of 6247 \AA\ during 68 epochs, and the IO:O CCD camera (Sloan 
$r$ filter with CWL = 6187 \AA) during 145 epochs. The properties of both cameras, 
exposure times, and pre-processing tasks were described in \citetalias{Goic16}. For 
the LT monitoring campaign between 2009 and 2019, the median value of the full-width 
at half-maximum (FWHM) of the seeing disc was $1\farcs40$. The first part of the 
Panoramic Survey Telescope and Rapid Response System project \citep[PS1;][]{Cham19}  
also imaged the double quasar in the $r$ band with CWL = 6215 \AA. The PS1 Data 
Release 2\footnote{\url{http://panstarrs.stsci.edu}} \citep{Flew20} included warp 
frames at six epochs over the 2011$-$2014 period. Despite the use of short exposures 
of only 40 s, the pixel scale of $0\farcs258$ pixel$^{-1}$ and good seeing conditions 
(median FWHM of $1\farcs07$) yielded useful photometric data (see below).

\setcounter{table}{1}
\begin{table*}
\centering
\caption{Spectroscopic observations in the 2009$-$2019 period.}
\begin{tabular}{lccccc}
\hline\hline
Date & Instrumentation\tablefootmark{a} &
Wavelength range (\AA) & Res. Power & Main emission lines & Ref \\
\hline
2012-Mar-16 & SDSS-BOSS          & 3600$-$10400  &    2000        &
Ly$\alpha$, \ion{C}{iv}, \ion{C}{iii}], \ion{Mg}{ii}           & 1 \\
2012-Mar-29 & SDSS-BOSS          & 3600$-$10400  &    2000        &
Ly$\alpha$, \ion{C}{iv}, \ion{C}{iii}], \ion{Mg}{ii}           & 1 \\
2013-Apr-13 & GTC-OSIRIS-R500R   & 4850$-$9250   &     400        &
\ion{C}{iv}, \ion{C}{iii}], \ion{Mg}{ii}                       & 2 \\
2014-Mar-27 & GTC-OSIRIS-R500R   & 4850$-$9250   &     400        &
\ion{C}{iv}, \ion{C}{iii}], \ion{Mg}{ii}                       & 3 \\
2014-May-20 & GTC-OSIRIS-R500B   & 3600$-$7200   &     320        &
Ly$\alpha$, \ion{C}{iv}, \ion{C}{iii}]                         & 3 \\
2016-Feb-18 & HST-WFC3-UVIS      & 2000$-$6000   &      70        &
Ly$\alpha$, \ion{C}{iv}                                        & 4 \\
2017-Apr-06 & VLT-XSHOOTER       & 3050$-$20700  & 6700, 8900, 5600 &
Ly$\alpha$, \ion{C}{iv}, \ion{C}{iii}], \ion{Mg}{ii}, H$\beta$, [\ion{O}{iii}] & 5 \\
2018-Feb-23 & VLT-XSHOOTER       & 3050$-$20700  & 6700, 8900, 5600 &
Ly$\alpha$, \ion{C}{iv}, \ion{C}{iii}], \ion{Mg}{ii}, H$\beta$, [\ion{O}{iii}] & 5 \\
2019-Mar-15 & TNG-NICS-HK        & 13500$-$24700 &     500        &
H$\alpha$                                                      & 6 \\
2019-Jun-10 & NOT-ALFOSC-G18     & 3450$-$5350   &    1000        &
Ly$\alpha$, \ion{C}{iv}                                        & 6 \\
\hline
\end{tabular}
\tablefoot{
\tablefoottext{a}{SDSS-BOSS $\equiv$ Baryon Oscillation Spectroscopic Survey (BOSS) 
spectrograph for the SDSS, GTC-OSIRIS-R500x $\equiv$ OSIRIS instrument (R500x grism) 
on the 10.4 m Gran Telescopio Canarias (GTC), HST-WFC3-UVIS $\equiv$ UVIS channel of 
the WFC3 instrument on the {\it Hubble} Space Telescope (HST), VLT-XSHOOTER $\equiv$ 
XSHOOTER spectrograph (UVB, VIS, and NIR arms) on the 8.2 m Very Large Telescope (VLT), 
TNG-NICS-HK $\equiv$ NICS instrument (HK grism) on the 3.6 m Telescopio Nazionale 
Galileo (TNG), and NOT-ALFOSC-G18 $\equiv$ ALFOSC instrument (grism \#18) on the 2.5 m 
Nordic Optical Telescope (NOT)}. \\
}
\tablebib{
(1) \citet{Daws13}; (2) \citetalias{Shal14}; (3) \citetalias{Goic16}; (4) 
\citet{Luss18}; (5) VLT-XSHOOTER programme 099.A-0018 (PI: M. Fumagalli); (6) This 
paper.
}
\label{tab:spec}
\end{table*}

We also used frames of the lens system taken with the 1.5 m Maidanak Telescope (MT) on 
six nights in July and December 2010. These observations were carried out with the 
SNUCAM camera and the Bessell $R$ filter \citep[CWL = 6462 \AA;][]{Im10}. SNUCAM uses 
a CCD detector with a $0\farcs266$ pixel$^{-1}$ scale, and several 180 or 300 s 
exposures were obtained at each of the six epochs. Basic pre-processing tasks were 
then applied to MT frames: bias subtraction, trimming, flat fielding, and World 
Coordinate System mapping. The median FWHM was $1\farcs15$. In addition, the science 
archive of the National Optical Astronomy 
Observatory\footnote{\url{http://archive.noao.edu}} provides public access to frames 
of \object{SDSS J1339+1310} taken with the ANDICAM instrument \citep{DePo03} on the 
1.3 m SMARTS Telescope (ST). We selected the $R$-band exposures (CWL = 6576 \AA) at 16 
epochs over the 2013$-$2016 period. ANDICAM has a $0\farcs37$ pixel$^{-1}$ scale, and 
three 300 s exposures are available for each observation night. These ST frames were 
conveniently pre-processed before starting photometry. The median FWHM for the 16 
observing epochs was $1\farcs50$. 

Our photometric technique is detailed in \citetalias{Goic16}, but we give a brief 
outline of main steps here. We obtained quasar fluxes through point-spread function 
(PSF) fitting as well as the IRAF \citep{Tody86,Tody93} and IMFITFITS \citep{McLe98} 
software programs. The 
photometric model consisted of two PSFs (images A and B), a de Vaucouleurs profile 
convolved with the PSF (light distribution of G) and a constant sky background, and 
it was applied to all LT, PS1, MT, and ST frames. The Sloan Digital Sky Survey (SDSS)
$r$-band magnitude of a reference star was used for calibrating quasar magnitudes, and 
typical errors of new LT data were estimated from root-mean-square deviations between 
magnitudes on consecutive nights (see \citetalias{Goic16}). MT errors were derived 
from standard deviations of magnitudes on each observation night, while we adopted the 
typical errors in LT records as the PS1 and ST photometric uncertainties. The 11-year 
light curves are available in Table 1 at the CDS: Column 1 lists the observing date 
(MJD$-$50\,000), Cols. 2$-$3 give magnitudes and magnitude errors of A, and Cols. 
4$-$5 give magnitudes and magnitude errors of B. These curves are also displayed in 
Figure~\ref{fig:lcur}. Although we mostly use observations in standard $r$ bands (in 
219 out of 241 epochs), there are data at slightly redder wavelengths in 22 epochs. 
This results in an effective, weighted average wavelength of 6237 \AA, which 
translates to a UV continuum emission at 1930 \AA. We also note that the LT data are in 
reasonably good agreement with those from PS1 and ST. 

The light curves over the full observing period 2009$-$2019 were initially used 
to cross-check the time delay measured in \citetalias{Goic16}. In 
Appendix~\ref{sec:appena}, we confirm the delay based on a shorter monitoring period 
and show that the adopted error bar is very conservative.

\subsection{UV-visible-NIR spectra}
\label{sec:spec}

Spectroscopic observations of \object{SDSS J1339+1310} were also performed between 
2009 and 2019. Main details are incorporated into Table~\ref{tab:spec}, and observing 
epochs are indicated in Figure~\ref{fig:lcur}. The calibrated SDSS-BOSS spectra are 
publicly available in the SDSS database\footnote{\url{https://www.sdss.org}}. However, 
the BOSS spectrograph collected light into a 2\arcsec-diameter optical fibre 
\citep{Daws13}, which was centred on image A in the first epoch and on image B in the 
second. Hence, the two-epoch observations do not provide clean spectra of A or B. 
For instance, in the first epoch, although the spectral energy distribution mainly 
corresponds to A, a significant contribution of B and the lensing galaxy is also expected. 
We also conducted long-slit spectroscopy with the GTC, the TNG, and the NOT as 
part of the GLENDAMA project. The GTC-OSIRIS observations allowed us to resolve A, B, 
and G, and accurately extract their individual spectra \citepalias{Shal14,Goic16}. 
Using the R500R grism, the C\,{\sc iv} emission line in quasar spectra appears close 
to its blue edge, whereas the Mg\,{\sc ii} emission line is located on its red edge. 
The C\,{\sc iv} emission of the quasar is, however, seen in the central part of the 
R500B grism wavelength range. Additionally, our spectroscopic follow-up of both quasar 
images with NOT-ALFOSC and TNG-NICS, provided relatively noisy shapes for the C\,{\sc 
iv} emission line and the first detection of H$\alpha$ emission. These recent spectra 
will be presented in a future paper and not further considered here.

\begin{figure}
\centering
\includegraphics[width=9cm]{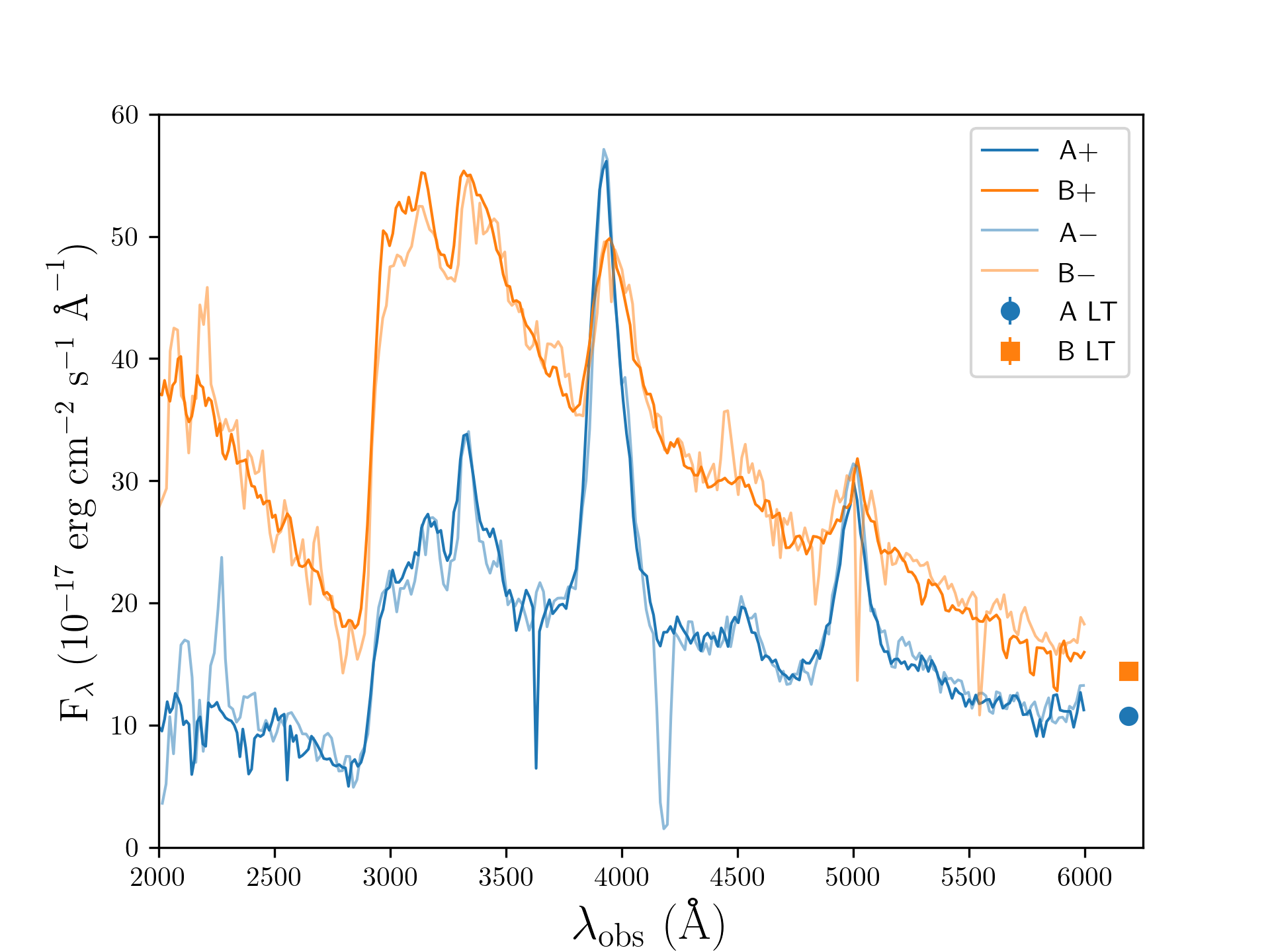}
\caption{HST-WFC3-UVIS spectra of SDSS J1339+1310AB in February 2016. The first order 
spectra are A$^{+}$ and B$^{+}$ (+1st order data), as well as A$^{-}$ and B$^{-}$ 
($-$1st order data). Fluxes from LT $r$-band frames on 14 February 2016 are also shown 
for comparison purposes.}
\label{fig:hstspec}
\end{figure}

The HST Data Archive\footnote{\url{http://archive.stsci.edu/hst}} also contains 
slitless spectroscopy of \object{SDSS J1339+1310} with the WFC3 instrument. The
observations were made using the G280 grism in the UVIS channel, and were presented by  
\citet{Luss18}. Unfortunately, \citet{Luss18} only displayed resulting spectra for the 
two quasar images in their Fig. 2. Thus, we downloaded the HST-WFC3-UVIS original data 
and then extracted the quasar spectra that are plotted in Figure~\ref{fig:hstspec}. 
The overall shape of +1st order spectra (labelled as A$^{+}$ and B$^{+}$) agrees with 
that of $-$1st order ones (A$^{-}$ and B$^{-}$). However, the throughput of the G280 
grism is higher for the +1st order, and the associated spectra are less noisy. We 
accordingly take A$^{+}$ and B$^{+}$ as our final data. These HST-WFC3-UVIS spectra 
are available in tabular format at the CDS: Table 3a includes wavelengths in \AA\ 
(Col. 1), and fluxes and flux errors of A in 10$^{-17}$ erg cm$^{-2}$ s$^{-1}$ 
\AA$^{-1}$ (Cols. 2 and 3), while Table 3b includes wavelengths in \AA\ (Col. 1), and 
fluxes and flux errors of B in 10$^{-17}$ erg cm$^{-2}$ s$^{-1}$ \AA$^{-1}$ (Cols. 2 
and 3). The spectral energy distributions show the \ion{C}{iv} emission line around 
5000 \AA\ ($\sim$1550 \AA\ in the quasar rest-frame).

\begin{figure}
\centering
\includegraphics[width=9cm]{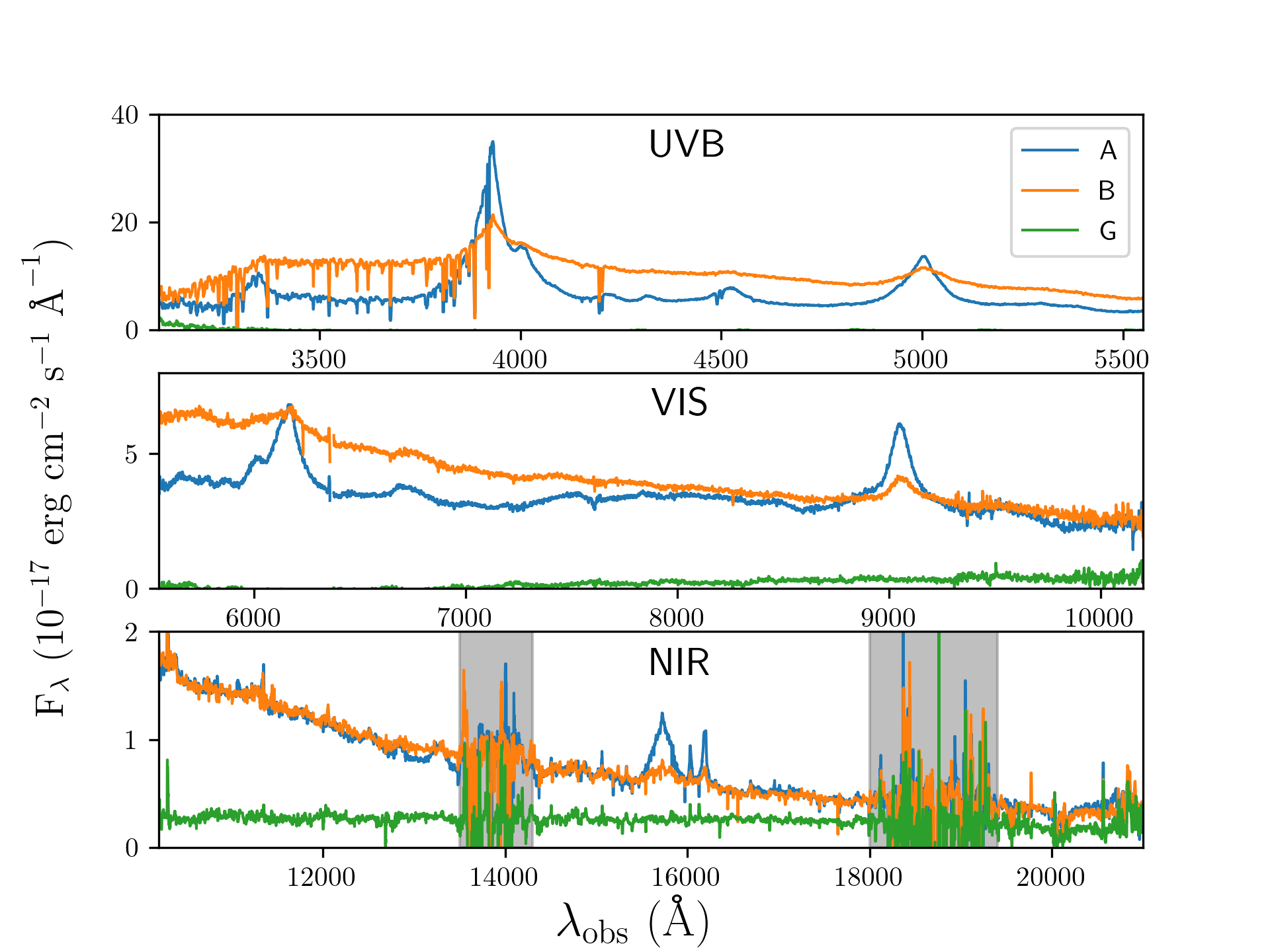}
\caption{VLT-XSHOOTER spectra of SDSS J1339+1310ABG in April 2017. The fluxes in the 
UVB, VIS, and NIR arms are depicted in the top, middle, and bottom panels, 
respectively. Grey highlighted regions display very noisy behaviours that are related 
to strong atmospheric absorption.}
\label{fig:vltspec}
\end{figure}

We also downloaded and analysed observations in the three arms (UVB, VIS, and NIR) of 
VLT-XSHOOTER \citep{Vern11}. These are publicly available at the ESO Science Archive 
Facility\footnote{\url{http://archive.eso.org/cms.html}}, and details on the observing 
programme are given in Table~\ref{tab:spec}. For each of the two epochs, we used the 
2700 (3$\times$900) s exposure and a three-component fitting to resolve the spectra of 
A, B, and G. Our multi-component extraction technique is robust and has been recently 
applied to some lens systems \citep[e.g.][]{Slus07,Shal17,Goic19}. The model we used 
can be described in a nutshell as follows: three 1D Moffat profiles in the spatial 
direction for each wavelength bin. Telluric absorption in the VIS and NIR arms was 
also corrected via the Molecfit software \citep{Smet15,Kaus15}. Final calibrated 
spectra of A, B, and G on 6 April 2017 are available at the CDS: Table 4 contains 
wavelengths in \AA\ (Col. 1), fluxes and flux errors of A (Cols. 2 and 3), fluxes and 
flux errors of B (Cols. 4 and 5), and fluxes and flux errors of G (Cols. 6 and 7). The 
wavelength coverage ranges from 3050$-$20700 \AA\ (UVB, VIS, and NIR arms), and the 
fluxes and their errors are expressed in units of 10$^{-17}$ erg cm$^{-2}$ s$^{-1}$ 
\AA$^{-1}$. These VLT-XSHOOTER spectra are also shown in Fig.~\ref{fig:vltspec}. In 
addition to \ion{C}{iv} emission of the quasar in the UVB arm spectra (top panel), 
Mg\,{\sc ii} and H$\beta$ emissions are also seen around 9000 \AA\ ($\sim$2800 \AA\ in 
the quasar rest-frame; middle panel) and 15700 \AA\ ($\sim$4860 \AA\ in the quasar 
rest-frame; bottom panel), respectively.  

\section{Extinction and lens models}
\label{sec:extlens}

The position on the sky of \object{SDSS J1339+1310} was used to estimate its 
extinction in the Milky Way at three wavelengths of interest. Based on the results of 
\citet{Schl11}, the NASA/IPAC Extragalactic 
Database\footnote{\url{https://ned.ipac.caltech.edu}} provided Milky Way transmission 
factors ($\epsilon_{\rm{MW}}$) of 0.93 at 4362 \AA, 0.98 at 9693 \AA, and 0.99 at 
16478 \AA. These wavelengths correspond to 1350, 3000, and 5100 \AA\ in the quasar 
rest-frame (see Sect.~\ref{sec:bhmass}). While the Galaxy produces equal extinction in 
both quasar images, the lensing galaxy gives rise to a differential extinction between 
images. According to \citetalias{Goic16}, emission lines in GTC-OSIRIS spectra and a 
linear extinction law in G \citep[e.g.][]{Prev84} lead to a dust extinction ratio 
(ratio between the transmission of B and that of A) of 1.33 $\pm$ 0.11 at 5500 \AA\ in 
the lens rest-frame. Additionally, GTC-OSIRIS and VLT-XSHOOTER spectra indicate that 
Mg\,{\sc ii} absorption at the lens redshift is much stronger in A than in B. 
Therefore, it is reasonable to assume that the lens galaxy's dust essentially affects 
image A, which translates to transmission factors ($\epsilon_{\rm{G,A}}$) of 0.56 
$\pm$ 0.09, 0.77 $\pm$ 0.06, and 0.86 $\pm$ 0.04 for light emission at 1350, 3000, and 
5100 \AA, respectively (adopting an achromatic transmission factor 
$\epsilon_{\rm{G,B}}$ = 1 for image B).

We also modelled the (macro)lensing mass using observational constraints in previous 
papers. We took the image positions, and the galaxy position, effective radius, 
ellipticity, and position angle in the fifth column of Table 1 of \citetalias{Shal14} 
as constraints. The image fluxes were also considered to derive lens models. These 
fluxes are consistent with the macrolens magnification ratio measured in 
\citetalias{Goic16} (see Sect.~\ref{sec:intro}). The overall set of constraints 
allowed us to fit 11 free parameters with d.o.f. = 0, where 'd.o.f.' denotes the 
degrees of freedom. We used the GRAVLENS/LENSMODEL software \citep{Keet01}, 
adopting a gravitational lens scenario that consists of three components. De 
Vaucouleurs (DV) and Navarro-Frenk-White \citep[NFW; e.g.][]{Nava96,Nava97} mass 
profiles describe the stellar (light traces mass) and dark components of the main 
deflector G, whereas an external shear (ES) accounts for additional deflectors. 

\setcounter{table}{4}
\begin{table*}
\centering
\caption{DV+NFW+ES lens models.}
\begin{tabular}{lcccccccccccc}
\hline\hline
$f_*$ & \multicolumn{2}{c}{$\kappa$} & & \multicolumn{2}{c}{$\kappa_{*}/\kappa$} & &
\multicolumn{2}{c}{$\gamma$} & & \multicolumn{2}{c}{$\mu$} & $\Delta t_{\rm{AB}}$ (d) \\
 & A & B & & A & B & & A & B & & A & B & \\
\hline  
0.1 & 0.67 & 0.90 & & 0.02 & 0.05 & & 0.30 & 0.35 & & 52.1 & 9.1 & 22.7 \\
0.2 & 0.62 & 0.86 & & 0.05 & 0.11 & & 0.34 & 0.45 & & 31.7 & 5.5 & 27.7 \\
0.3 & 0.73 & 0.84 & & 0.07 & 0.17 & & 0.23 & 0.40 & & 42.7 & 7.5 & 21.5 \\
0.4 & 0.65 & 0.79 & & 0.10 & 0.24 & & 0.29 & 0.52 & & 25.1 & 4.4 & 28.1 \\
0.5 & 0.57 & 0.73 & & 0.14 & 0.32 & & 0.36 & 0.65 & & 16.3 & 2.9 & 34.6 \\
0.6 & 0.48 & 0.68 & & 0.20 & 0.42 & & 0.42 & 0.77 & & 11.5 & 2.0 & 41.1 \\
0.7 & 0.40 & 0.63 & & 0.28 & 0.52 & & 0.49 & 0.90 & &  8.5 & 1.5 & 47.6 \\
0.8 & 0.32 & 0.58 & & 0.40 & 0.65 & & 0.56 & 1.02 & &  6.6 & 1.2 & 54.2 \\
0.9 & 0.24 & 0.52 & & 0.60 & 0.81 & & 0.62 & 1.15 & &  5.2 & 0.9 & 60.7 \\
1.0 & 0.16 & 0.47 & & 1.00 & 1.00 & & 0.69 & 1.27 & &  4.3 & 0.7 & 67.2 \\
\hline
\end{tabular}
\tablefoot{
The ten lens models fit the observational constraints with $\chi^2 \sim$ 0 
(d.o.f. = 0; see main text). The parameter $f_*$ represents the luminous mass of G 
relative to that for the model without dark matter halo ($f_*$ = 1). We list the 
convergence ($\kappa$), stellar to total convergence ratio ($\kappa_{*}/\kappa$), 
shear ($\gamma$), and macrolens magnification ($\mu$) for each model and image 
position. The last column shows the predicted time delay between images for each 
model.
}
\label{tab:lensmod}
\end{table*}

The sequence of DV+NFW+ES models started by assuming that all mass of G is traced by 
light, namely constant mass-to-light ratio model. The free parameters of such a model 
were the position, mass scale, effective radius, ellipticity, and position angle of 
the DV profile, the strength and direction of the ES, and the position and flux of the 
source quasar. By progressively decreasing the luminous component of G and adding a 
concentric dark matter halo, we completed a sequence of ten realistic lens models with 
$f_*$ ranging between 1.0 and 0.1, where $f_*$ is the mass of G in stars (DV 
profile) relative to its maximum value in the absence of a dark matter halo \citep[we 
note that $f_*$ was called $f_{M/L}$ in several previous papers; 
e.g.][]{Morg18,Corn20}. For the $f_* <$ 1 models, the free parameters were the common 
position, ellipticity, and position angle for both components of G (DV and NFW 
profiles), the effective radius of the DV profile, the mass scale of the NFW profile, 
the strength and direction of the ES, and the position and flux of the source quasar. 

Some local parameters (at image positions) of the DV+NFW+ES models are shown in 
Table~\ref{tab:lensmod}. The last three columns of Table~\ref{tab:lensmod} give the 
magnification factor of A and B for each time delay, so we took the measured delay 
\citepalias[$\Delta t_{\rm{AB}}$ = 47$^{+5}_{-6}$ d;][and discussion in 
Appendix~\ref{sec:appena}]{Goic16} as an additional constraint to estimate 
reliable intervals for $\mu_{\rm{A}}$ and $\mu_{\rm{B}}$. First, results for the 
models in the range 0.6 $\leq f_* \leq$ 0.8 were used to construct power-law 
cross-correlations between magnifications and delays. If $y = \mu_{\rm{A}}$ (or 
$\mu_{\rm{B}}$) and $x = \Delta t_{\rm{AB}}$, we derived laws $y \propto x^{-\alpha}$, 
where the power-law index is $\alpha \approx$ 2. Second, these $x$-$y$ relationships 
and the extreme values of the measured delay interval ($x$ = 41 and 52 d) led to 
$\mu_{\rm{A}}$ = 9.4 $\pm$ 2.2 and $\mu_{\rm{B}}$ = 1.6 $\pm$ 0.4. 

\section{Black hole mass}
\label{sec:bhmass}

For a given broad emission line, it is thought that the line-emitting gas is 
distributed within the gravitational potential of the central SMBH, so there is a 
relationship between its motion, the size of the region, and the SMBH mass 
\citep[$M_{\rm{BH}}$; e.g.][and references therein]{Vest06}. The gas motion is 
responsible for the emission-line width, while the radius of the line-emitting region 
scales roughly as the square root of the continuum luminosity 
\citep[e.g.][]{Kora91,Bent09}. Hence, we aimed to use the single-epoch spectra of 
\object{SDSS J1339+1310} in Table~\ref{tab:spec} to measure the black hole mass of the 
quasar. Although there are spectra for the two quasar images, image B is strongly 
affected by microlensing. Thus, we only considered image A when estimating 
$M_{\rm{BH}}$. We were primarily interested in the width of the \ion{C}{iv}, Mg\,{\sc 
ii}, and H$\beta$ emission lines, as well as the continuum flux for emissions at 1350, 
3000, and 5100 \AA\ \citep[e.g.][]{Vest06,Vest09}. After the de-redshifting of the observed
spectra of A to the quasar rest-frame, $\lambda_{\rm{rest}} = \lambda_{\rm{obs}} / (1 
+ z_{\rm{s}})$, if the continuum flux $F_{\rm{cont},\rm{A}}(\lambda_{\rm{rest}})$ is 
in 10$^{-17}$ erg cm$^{-2}$ s$^{-1}$ \AA$^{-1}$ and the corresponding luminosity 
$L_{\rm{cont}}(\lambda_{\rm{rest}})$ is in erg s$^{-1}$, then both quantities are 
related through the equation
\begin{equation}
    L_{\rm{cont}}(\lambda_{\rm{rest}}) = \frac{1.22 \times 10^{42} \lambda_{\rm{rest}} 
    F_{\rm{cont},\rm{A}}(\lambda_{\rm{rest}})}{\epsilon_{\rm{MW}}(\lambda_{\rm{rest}}) 
    \epsilon_{\rm{G,A}}(\lambda_{\rm{rest}}) \mu_{\rm{A}}} \,.
 \label{eq1}
 \end{equation}
All factors in the denominator of Eq.~(\ref{eq1}) are discussed in 
Sect.~\ref{sec:extlens}. 

\begin{figure}
\centering
\includegraphics[width=9cm]{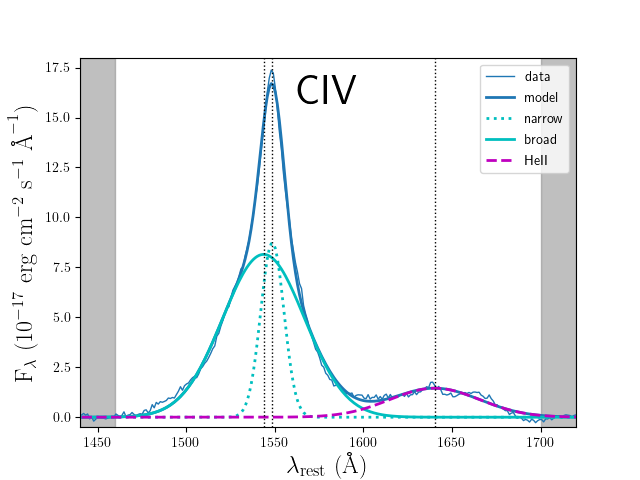}
\includegraphics[width=9cm]{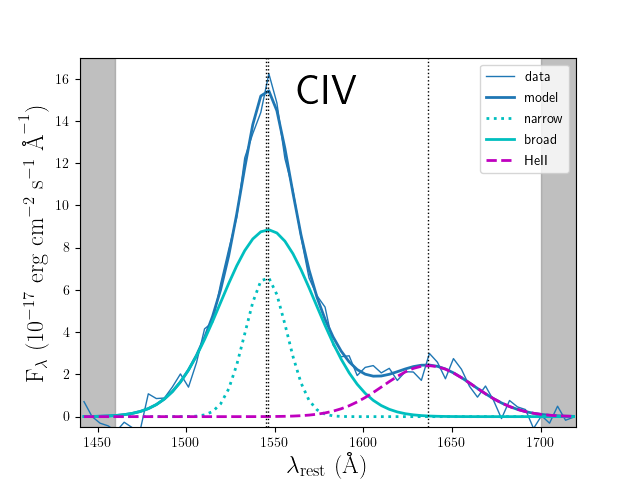}
\caption{Multi-component decomposition of the \ion{C}{iv} line profile in SDSS 
J1339+1310A. The profile is decomposed into three Gaussian components: \ion{C}{iv} 
broad + \ion{C}{iv} narrow + \ion{He}{ii} complex. The grey rectangles highlight the 
two spectral regions that we used to remove a linear continuum under the emission 
line, and the vertical dotted lines correspond to the centres of the Gaussians. The 
{\it top panel} displays GTC-OSIRIS-R500B data in May 2014, while the {\it bottom 
panel} incorporates HST-WFC3-UVIS data in February 2016.}
\label{fig:CIV}
\end{figure}

\begin{figure}
\centering
\includegraphics[width=9cm]{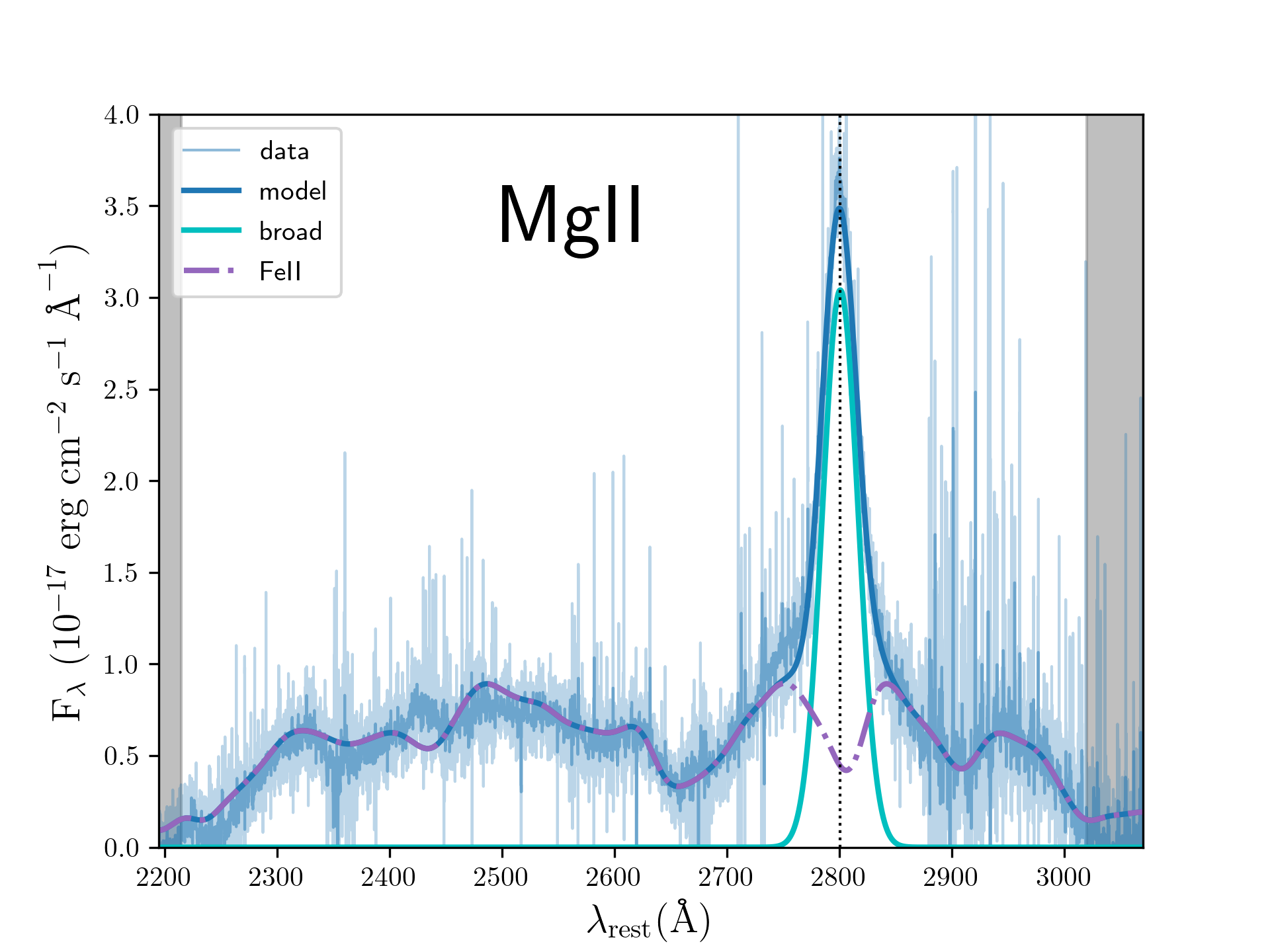}
\includegraphics[width=9cm]{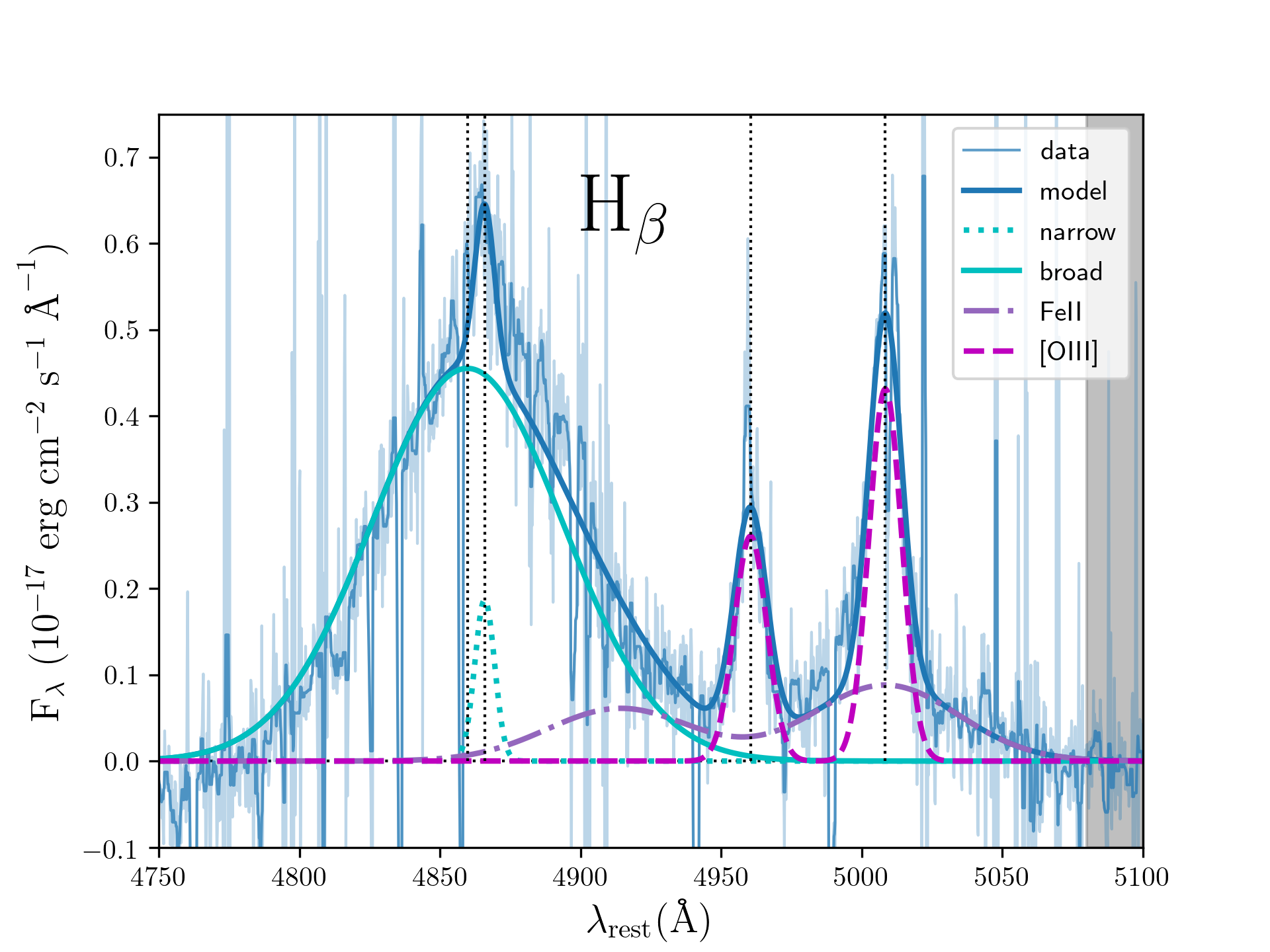}
\caption{Multi-component decomposition of the Mg\,{\sc ii} and H$\beta$ line profiles 
in SDSS J1339+1310A. The data correspond to VLT-XSHOOTER observations in April 2017. 
The original wavelength bins contain a very noisy signal (lighter blue), so a 
nine-point median filter is used to smooth original data (darker blue). In the {\it 
top panel}, after subtracting a power-law continuum, the residual signal is modelled 
as a sum of two contributions, i.e. Mg\,{\sc ii} broad emission (Gaussian curve around 
the vertical dotted line) and Fe\,{\sc ii} pseudo-continuum. There is no evidence of a 
Mg\,{\sc ii} narrow line or an additional very broad component. The continuum windows 
at 2195$-$2215 \AA\ and 3020$-$3070 \AA\ are highlighted using grey rectangles, and 
the Fe\,{\sc ii} pseudo-continuum is created from the Fe\,{\sc ii} template of 
\citet{Tsuz06} convolved with a Gaussian function to account for the Doppler 
broadening of Fe\,{\sc ii} lines. The {\it bottom panel} displays the decomposition in 
the spectral region around the H$\beta$ emission. The grey rectangle highlights one of 
the two windows that we used to remove a power-law continuum under emission lines 
\citep[3790$-$3810 \AA\ and 5080$-$5100 \AA;][]{Kura02}. The residual signal is 
decomposed into four Gaussian components: H$\beta$ broad + H$\beta$ narrow + [O\,{\sc 
iii}] doublet, plus the Fe\,{\sc ii} contribution from two individual lines in the 
spectral range 4750$-$5100 \AA\ \citep[e.g.][]{Kova10}. Vertical dotted lines show the 
centres of the four Gaussians.}
\label{fig:MgIIHb}
\end{figure}

The \ion{C}{iv} emission line is well resolved in the GTC-OSIRIS-R500B spectrum of 
image A. In Fig. 4 of \citetalias{Goic16}, we presented a multi-component 
decomposition of its profile in both images. For clarity, the carbon line profile in A 
is also depicted in the top panel of Figure~\ref{fig:CIV}. After subtracting the local 
continuum, the residual signal is modelled as a sum of three Gaussian contributions. 
We focused on the \ion{C}{iv} total (broad + narrow) emission line and calculated the 
square root of its second moment \citep[$\sigma_{\rm{line}}$;][]{Pete04}. The line 
width was then estimated from the spectral resolution-corrected line dispersion 
$\sigma_{\rm{l}} = (\sigma_{\rm{line}}^2 - \sigma_{\rm{inst}}^2)^{1/2}$ = 3978 km 
s$^{-1}$. Moreover, the continuum flux at $\lambda_{\rm{rest}}$ = 1350 \AA\ in May 
2014 reached a level of 5.9 $\times$ 10$^{-17}$ erg cm$^{-2}$ s$^{-1}$ \AA$^{-1}$, 
that is $F_{\rm{cont},\rm{A}}(1350 \rm{\AA})$ = 5.9. We additionally considered the 
HST-WFC3-UVIS spectrum of A and analysed the region around the \ion{C}{iv} line (see 
the bottom panel of Figure~\ref{fig:CIV}). Thus, we obtained an independent line width 
$\sigma_{\rm{l}}$ = 4294 km s$^{-1}$ based on HST data. Despite the fact that both 
spectra (GTC and HST) yield similar values of $\sigma_{\rm{l}}$, the 
$F_{\rm{cont},\rm{A}}(1350 \rm{\AA})$ values in the two observing epochs were very 
different. The continuum flux in February 2016 was $F_{\rm{cont},\rm{A}}(1350 
\rm{\AA})$ = 15.6. A dramatic change in the $r$-band flux between May 2014 (minimum 
level) and February 2016 (maximum level) is also seen in Figure~\ref{fig:lcur}. From 
Eq.~(\ref{eq1}), accounting for uncertainties in the dust extinction and macrolens 
magnification of image A, we found ${\log [L_{\rm{cont}}(1350 \rm{\AA})]}$ = 45.30 
$\pm$ 0.13 (GTC) and 45.72 $\pm$ 0.13 (HST).

\citet{Vest06} derived equations for estimating the central black hole mass in a 
quasar from the \ion{C}{iv} line width and $L_{\rm{cont}}(1350 \rm{\AA})$, and we used 
their Eq. (8) to obtain \ion{C}{iv}-based masses. The spectral observations in May 
2014 (GTC) and February 2016 (HST) led to ${\log \left[ 
M_{\rm{BH}}(\rm{\ion{C}{iv}})/\rm{M_{\odot}} \right]}$ = 8.62 $\pm$ 0.33 and 8.91 
$\pm$ 0.33, respectively. We have not considered uncertainties generated by errors in 
$\sigma_{\rm{l}}$ and $L_{\rm{cont}}(1350 \rm{\AA})$ since the uncertainty in ${\log  
\left[ M_{\rm{BH}}(\rm{\ion{C}{iv}})/\rm{M_{\odot}} \right]}$ is dominated by an 
intrinsic scatter of $\pm$0.33 dex rather than contributions from such errors. An 
error of 100 km s$^{-1}$ in $\sigma_{\rm{l}}$ propagates to $\pm$0.02 dex in ${\log 
\left[ M_{\rm{BH}}(\rm{\ion{C}{iv}})/\rm{M_{\odot}} \right]}$, whereas the 
uncertainties in $L_{\rm{cont}}(1350 \rm{\AA})$ propagate to $\pm$0.07 dex in the 
logarithm of the mass. Typical mass logarithms of 8.56 (GTC) and 8.83 (HST) were also 
inferred from Eq. (2) of \citet{Park13}.

\begin{figure*}
\centering
\includegraphics[width=15cm]{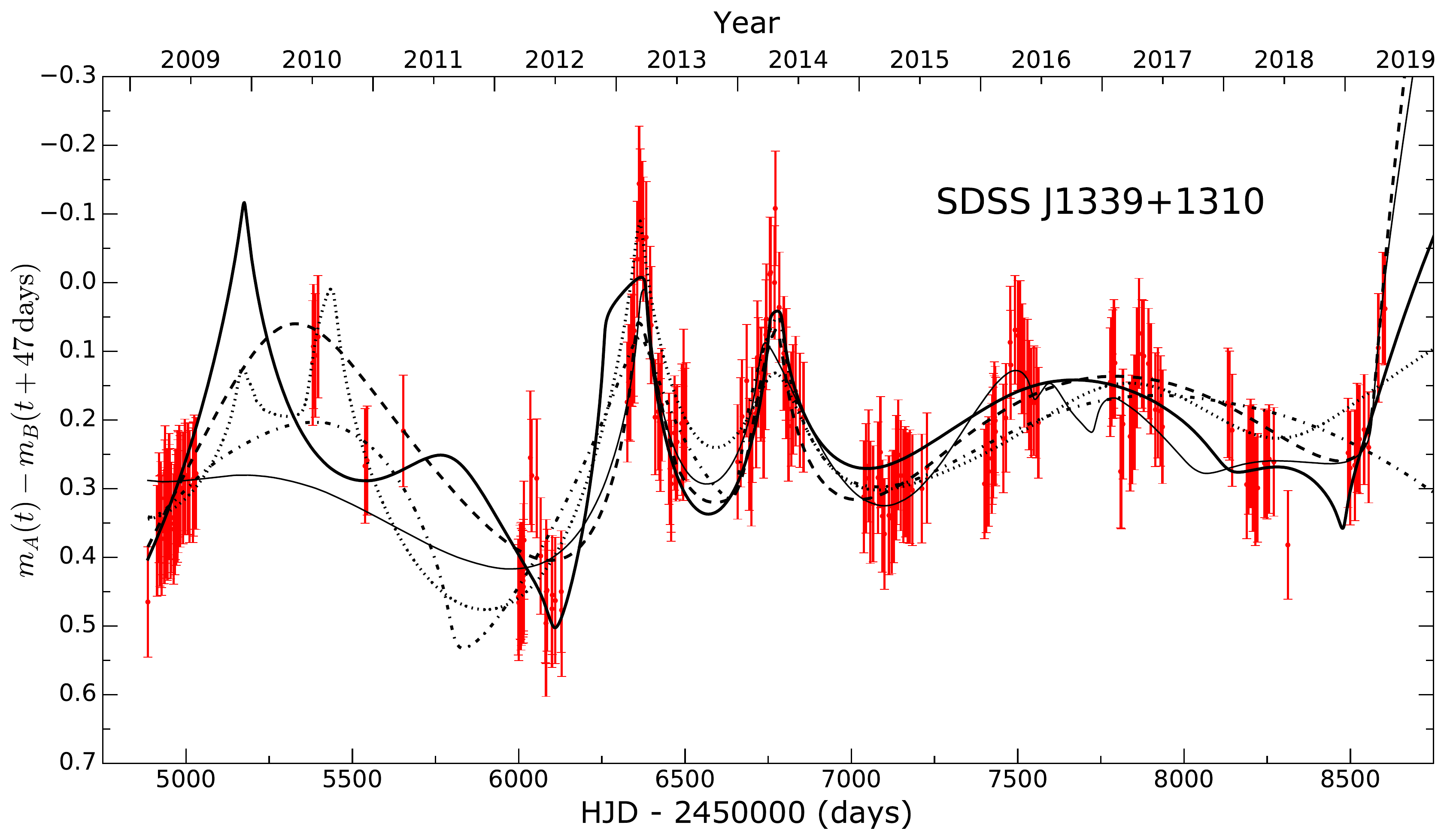}
\caption{Image B $r$-band light curve from the LT shifted by the 47-d time delay 
and subtracted from the contemporaneous magnitude of image A, i.e. $m_{\rm A}(t) - 
m_{\rm B}(t + {\rm 47~d})$ (equivalent to a ratio in flux units), leaving only 
variability that is not intrinsic to the source itself. Plotted in black are five good 
fits to the microlensing variability from our Monte Carlo routine.}
\label{fig:bestfits}
\end{figure*}

Although both \ion{C}{iv}-based masses are consistent within errors, and this would 
permit us to 'confidently' calculate the average, we carefully analysed the Mg\,{\sc 
ii} and H$\beta$ emission lines in the VLT-XSHOOTER spectrum of A in April 2017. These 
VLT observations were made at an epoch for which the $r$-band flux reached an 
intermediate level (see Figure~\ref{fig:lcur}), and therefore they may help to decide 
whether one of the two \ion{C}{iv}-based estimates is relatively biased or not. In 
Figure~\ref{fig:MgIIHb}, we show multi-component decompositions of the Mg\,{\sc ii} 
and H$\beta$ profiles (see the caption for details). First, we measured FWHM(Mg\,{\sc 
ii}) = 3984 km s$^{-1}$ and $F_{\rm{cont},\rm{A}}(3000 \rm{\AA})$ = 2.3. In addition, 
Eq.~(\ref{eq1}) provided the relevant continuum luminosity at 3000 \AA, namely ${\log 
[L_{\rm{cont}}(3000 \rm{\AA})]}$ = 45.08 $\pm$ 0.11. Using Eq. (1) of \citet{Vest09} 
at $\lambda_{\rm{rest}}$ = 3000 \AA, Eq. (10) of \citet{Wang09}, and Eq. (3) of 
\citet{Shen12}, we obtained typical ${\log \left[ 
M_{\rm{BH}}(\rm{\ion{Mg}{ii}})/\rm{M_{\odot}} \right]}$ values ranging between 8.58 
and 8.70, in good agreement with those from the C\,{\sc iv} line width and 
$L_{\rm{cont}}(1350 \rm{\AA})$ in the GTC spectrum taken in May 2014. 

From the decomposition in the bottom panel of Figure~\ref{fig:MgIIHb} and 
Eq.~(\ref{eq1}), we also derived FWHM(H$\beta$) = 4928 km s$^{-1}$, 
$F_{\rm{cont},\rm{A}}(5100 \rm{\AA})$ = 0.53, and ${\log [L_{\rm{cont}}(5100 
\rm{\AA})]}$ = 44.61 $\pm$ 0.11. Following the prescription of \citet{Vest06}, we only 
used the broad component of the H$\beta$ emission to determine FWHM(H$\beta$). The 
FWHM(H$\beta$) and ${\log [L_{\rm{cont}}(5100 \rm{\AA})]}$ values yielded ${\log 
\left[ M_{\rm{BH}}(\rm{H\beta})/\rm{M_{\odot}} \right]}$ = 8.60 $\pm$ 0.43 \citep[Eq. 
(5) of][]{Vest06}, indicating an excellent agreement between the new measurement and 
${\log \left[ M_{\rm{BH}}(\rm{\ion{C}{iv}})/\rm{M_{\odot}} \right]}$ from GTC data. We 
finally adopted ${\log \left( M_{\rm{BH}}/\rm{M_{\odot}} \right)}$ = 8.6 $\pm$ 0.4. 
This interval encompasses all our black-hole mass measurements from GTC and VLT 
spectra.

\section{Accretion disc size from microlensing variability}
\label{sec:microsize}

We analysed the $r$-band light curves using the Monte Carlo analysis technique in 
\citet{Koch04}. The method is fully described in that paper, but we provide a brief 
summary here. Using the convergence $\kappa$, convergence due to stars $\kappa_*$, 
shear $\gamma$, and shear position angle $\theta_{\gamma}$ from each of the models in 
our sequence (see Table~\ref{tab:lensmod}), we generated a set of 40 
magnification patterns per macro model to represent the microlensing conditions at the 
location of each lensed image. Our model sequence has 10 macro models in the range 
$0.1 \leq f_* \leq 1.0$ (see Sect.~\ref{sec:extlens}), so we generated a total of 400 
sets of magnification patterns.

Each of the 400 realisations was generated randomly using the inverse ray-shooting
technique as described by \citet{Koch06}, the distribution of stellar masses for which 
were drawn from the Galactic bulge IMF of \citet{Goul00} in which $dN(M)/dM\propto 
M^{-1.3}$. The square magnification patterns represent a region 40 $r_{\rm E}$ on a 
side, where $r_{\rm E}$ is the source-plane projection of the Einstein radius of a 
$1{\rm M_{\odot}}$ star. With dimensions of $8192 \times 8192$ pixels, each pixel 
represents a physical distance of $1.60 \times 10^{14} \avgmhat^{1/2}$~cm on the 
source plane, where $\avgmstar$ is the mean mass of a lens galaxy star.

\begin{figure*}
\centering
\includegraphics[width=9cm]{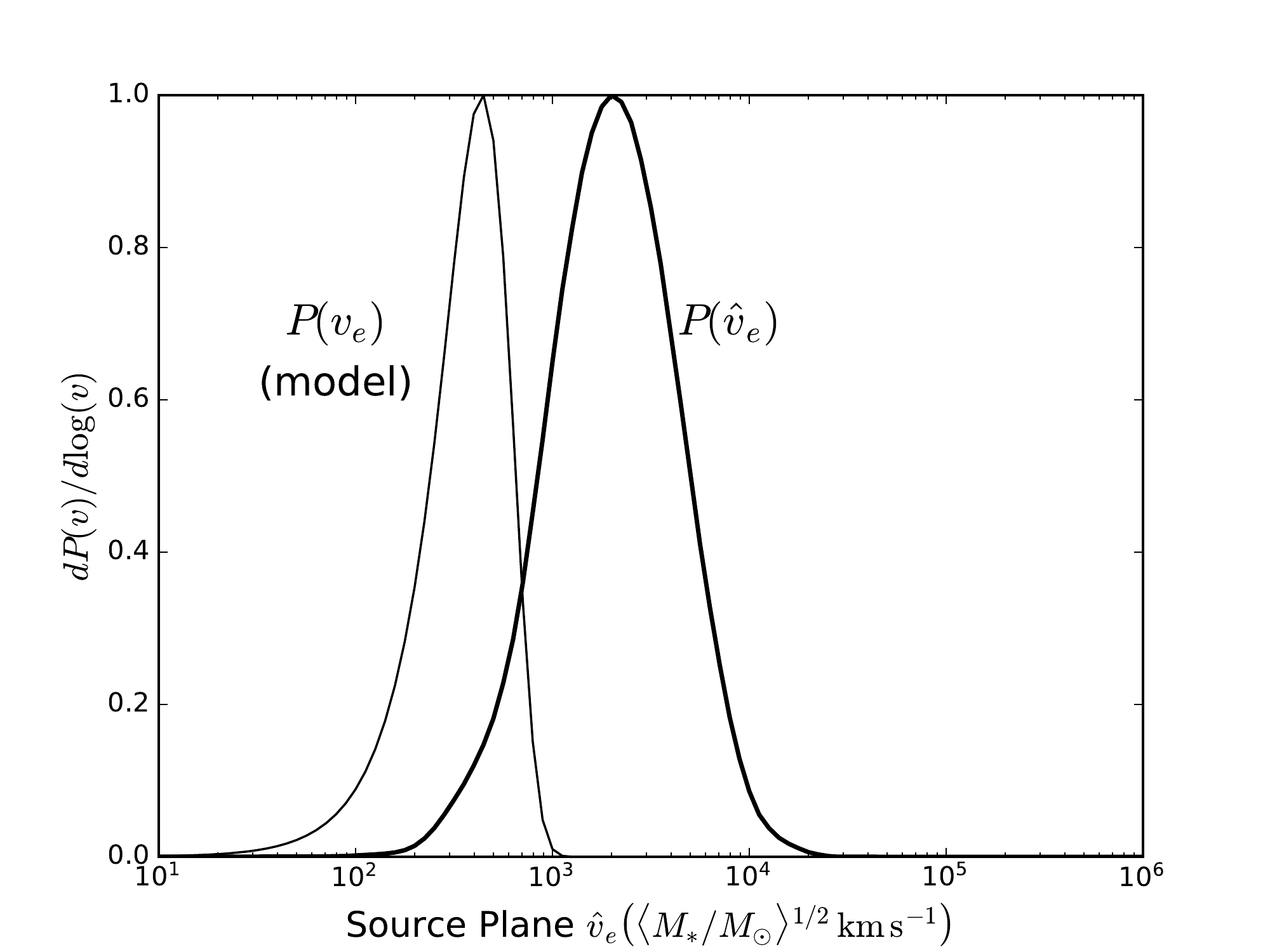}
\includegraphics[width=9cm]{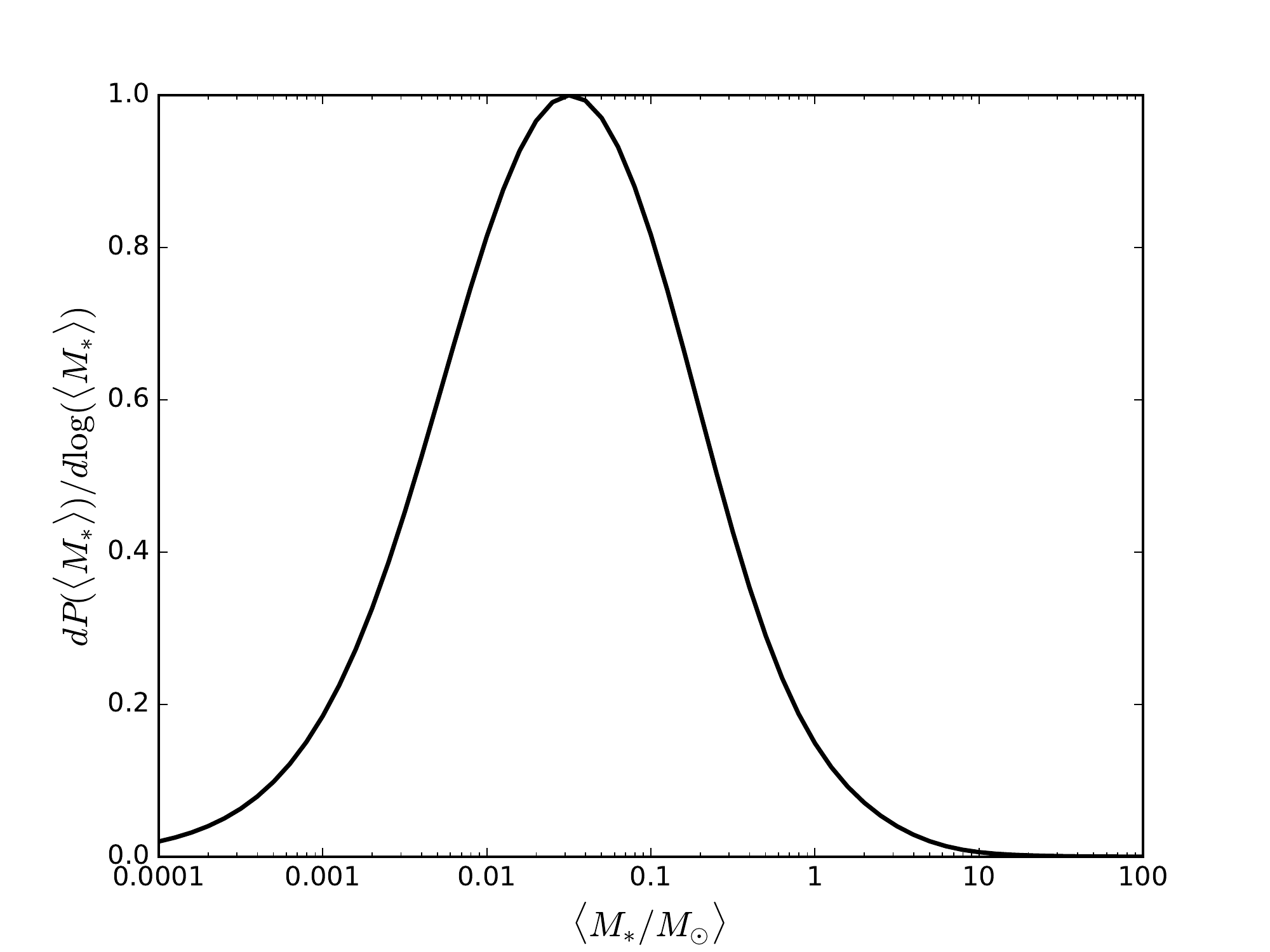}
\caption{Distributions for the effective source velocity and mean microlens mass. 
{\it Left panel}: Probability density for the effective velocity $\hat{v}_{\rm e}$ (in 
Einstein units $\avgmhat^{1/2}{\rm km \: s^{-1}}$) across the source plane from the 
ensemble of Monte Carlo fits to the observed light curves (heavy black curve). 
The lighter black curve is our prior on the true effective 
velocity of source (in physical units ${\rm km \: s^{-1}}$) generated from a model of 
the effective velocity of source, lens, observer and the velocity dispersion of lens 
galaxy stars. {\it Right panel}: Probability density for the mean mass of a lens 
galaxy star $\avgmstar$ derived by convolving the probability density for the 
effective velocity $\hat{v}_{\rm e}$ with our model for the true effective velocity 
(see left panel). Since $\avgmhat = (v_{\rm e} / \hat{v}_{\rm e})^2$, the 
distribution is broad. The expectation value for $\avgmstar = 0.029_{-0.025}^{+0.151} 
\: {\rm M_{\odot}}$ is lower than the predictions of standard IMF models, but our 
results for the accretion disc size are consistent when we assume a more physical 
uniform prior $0.1 \leq \avgmhat \leq 1.0$ (see the right panel of 
Fig.~\ref{fig:size}).}
\label{fig:velmass}
\end{figure*}

\begin{figure*}
\centering
\includegraphics[width=9cm]{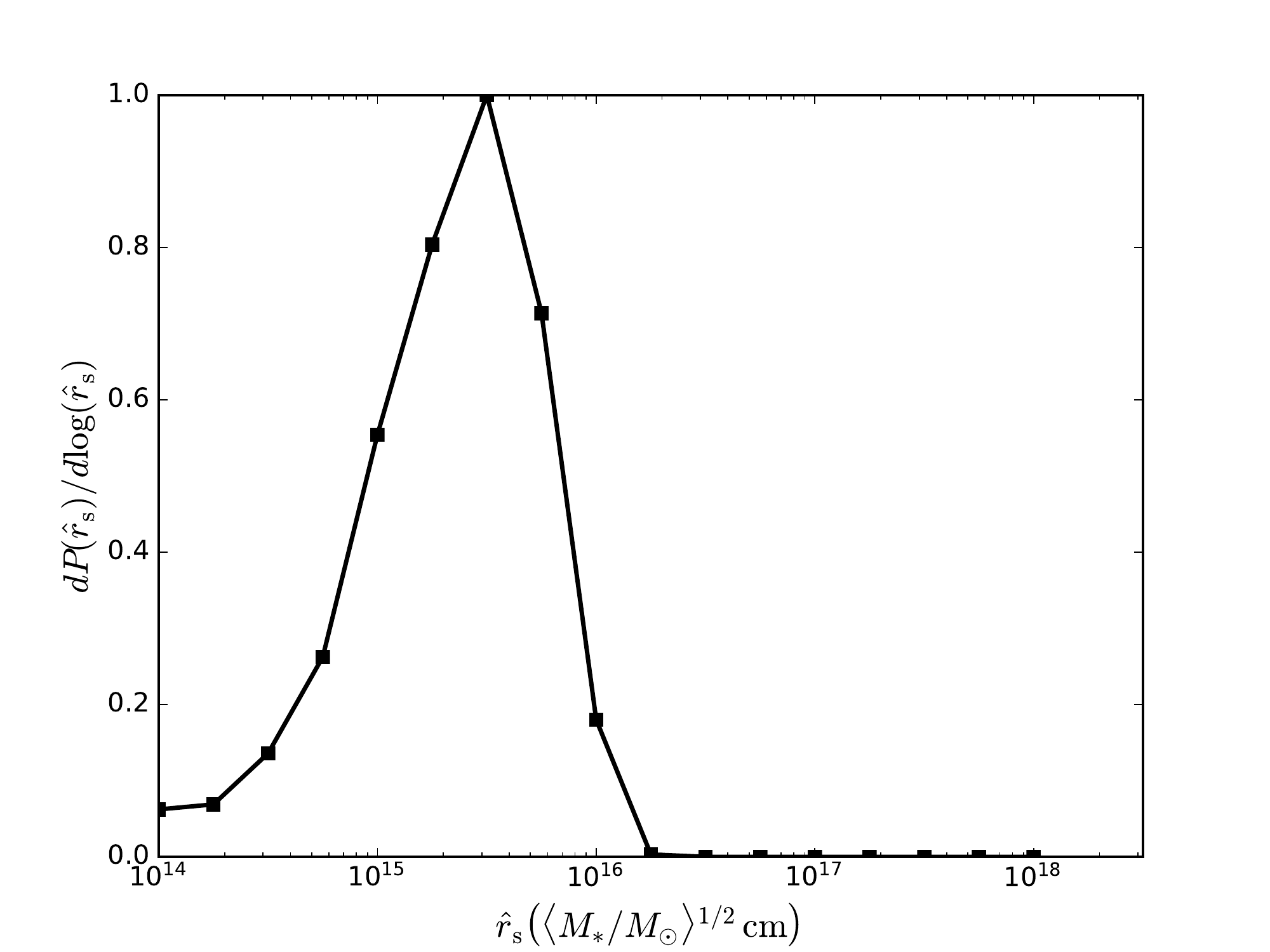}
\includegraphics[width=9cm]{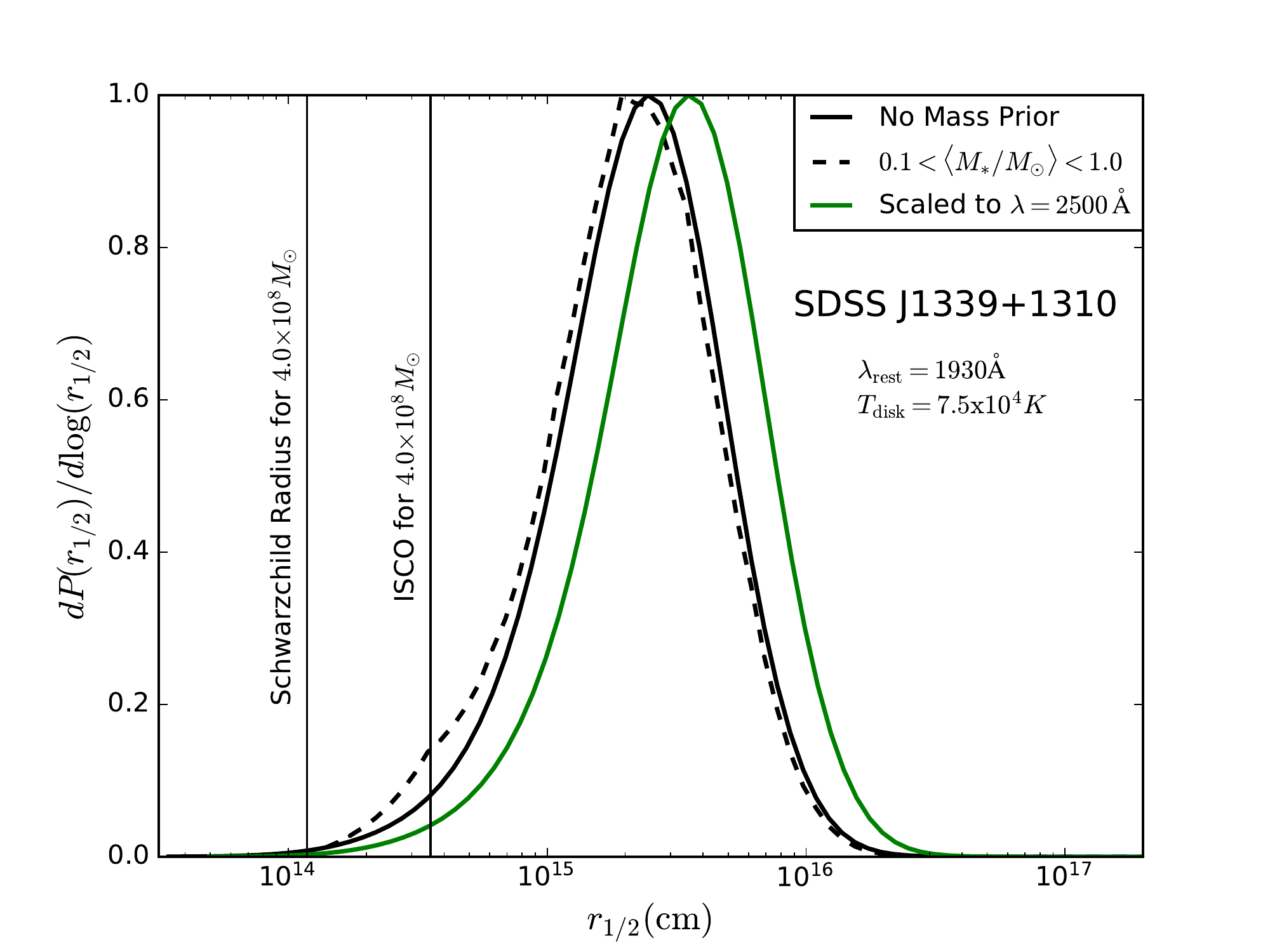}
\caption{Accretion disc structure in SDSS J1339+1310. 
{\it Left panel}: Probabilty density for the source size in Einstein units 
($\avgmhat^{1/2} \: {\rm cm}$). The distribution levels off at $\sim2\times10^{14} \: 
{\rm cm}$ since the pixel scale in our magnification patterns is $1.6\times10^{14} \: 
{\rm cm}$. All solutions become equally likely below that size scale. {\it Right 
panel}: Probability density for the half light radius of the continuum-emission region 
in SDSS J1339+1310 at $\lambda_{\rm rest}$ = 1930 \AA\ assuming an inclination angle 
$\cos i = 0.5$. The solid black curve was derived by convolving our probability 
density for $\avgmstar$ (right panel of Fig.~\ref{fig:velmass}) with that of 
$\hat{r}_s$ (see left panel). The dashed black curve assumed a uniform prior on the 
mean mass of a lens galaxy star $0.1 \leq \avgmstar \leq 1.0$. The solid green curve 
shows the result without the mass prior scaled to 2500 \AA\ for comparison to other 
models. The Schwarzschild radius $r_{\rm Sch} = 2 G M_{\rm BH} / c^2$ and ISCO at 
$3r_{\rm Sch}$ are plotted for reference.}
\label{fig:size}
\end{figure*} 

We convolved the magnification patterns with a simple \citet{Shak73} thin disc 
model for the surface brightness profile of the accretion disc at a range of source 
sizes $14.0 \leq \log(\hat{r}_{\rm s} / \avgmhat^{1/2}\,{\rm cm}) \leq 18.0$. The 
quantity $\hat{r}_{\rm s}$ is a thin disc scale radius where the hat indicates that 
the distance is being reported in 'Einstein units' in which quantities are scaled by a 
factor of the mean mass of a star in the lens galaxy. For example, converting a scale 
radius or velocity in Einstein units into a physical unit is accomplished using 
$r_{\rm s} = \hat{r}_{\rm s} \avgmhat^{1/2}$ and $v_{\rm e} = \hat{v}_{\rm e} 
\avgmhat^{1/2}$, respectively.   
   
In essence our technique is an attempt to reproduce the observed microlensing 
variability using a realistic model for conditions in the lens galaxy. We run our 
model magnification patterns by a range of model source sizes on a range of 
trajectories in an attempt to fit the observed data. According to Bayes' theorem, the 
likelihood of the set of physical $\xi_{\rm p}$ and trajectory $\xi_{\rm t}$ 
parameters given the data $D$ is
\begin{equation}
   P(\xi_{\rm p},\xi_{\rm t} \vert D) \propto P(D \vert \xi_{\rm p},\xi_{\rm 
   t})P(\xi_{\rm p})P(\xi_{\rm t}), 
\end{equation} 
where $P(\xi_{\rm t})$ and $P(\xi_{\rm p})$ are the prior probabilities for the 
trajectory and physical variables, respectively. In a given trial, we chose the 
velocity randomly from the uniform logarithmic prior $1.0 \leq \log[\hat{v}_{\rm e} / 
(\avgmhat^{1/2} \, {\rm km \, s^{-1}})] \leq 5.0$. We evaluated the goodness-of-fit 
with the chi-square statistic in real time during each trial, and for computational 
efficiency we aborted and discarded any fit with $\chi^2$/d.o.f. $\geq$ 1.6 since 
they do not contribute significantly to the Bayesian integrals. We attempted $10^7$ 
Monte Carlo fits per set of magnification patterns for a grand total of $4\times10^9$ 
trials. Of those trials, $\sim 10^5$ met our $\chi^2$ threshold, and the remainder of 
our analysis was performed on those solutions. In Figure~\ref{fig:bestfits}, we 
display examples of five good fits to the observed microlensing variability.  

We generated a statistical prior for the effective velocity between source, lens and 
observer using four components. We found the transverse velocity of the observer by 
calculating the northern ($v_{\rm O,n} = -110.7 \: {\rm km \, s^{-1}}$) and eastern ($v_{\rm 
O,e} = -218.5\: {\rm km \, s^{-1}}$) components of the CMB dipole across the line of 
sight to the target. Following \citet{Bolt08}, we estimated the velocity dispersion 
$\sigma_* = 379 \: {\rm km \, s^{-1}}$ in the lens galaxy using the monopole term of 
its gravitational potential. We estimated the peculiar velocity of source and lens 
using their redshifts, following the prescription of \citet{Mosq11}. Using the 
technique presented in \citet{Koch04}, we combined the velocities to generate a 
probability density for the effective velocity (see the left panel of 
Figure~\ref{fig:velmass}). Also shown in the left panel of Figure~\ref{fig:velmass} is 
the probability density for the effective velocity in Einstein units $\hat{v}_{\rm e}$ 
yielded by marginalising over the other variables in the set of successful solutions 
from the Monte Carlo run 
\begin{equation}
   P( \hat{v}_{\rm e} | D) \propto \int P(D| {\bf p}, \hat{v}_{\rm e} ) P({\bf p}) 
   P(\hat{v}_{\rm e}) d{\bf p},
\end{equation}
where $P(D| {\bf p}, \hat{v}_{\rm e})$ is the probability of fitting the data in a 
particular trial, $P({\bf p})$ sets the priors on the microlensing variables $\xi_{\rm 
p}$ \& $\xi_{\rm t}$, and $P(\hat{v}_{\rm e})$ is the (uniform) prior on the effective 
velocity. The total probability is then normalised so that $\int P(\hat{v}_{\rm e}|D)d 
\hat{v}_{\rm e}=1$. We carried out an analogous Bayesian integral to find the 
probability density for the source size in Einstein units, $d P(\hat{r}_{\rm s}) / d 
\log(\hat{r}_{\rm s})$, which we display in the left panel of Figure~\ref{fig:size}.

Now, to convert our result from Einstein units to true physical units, we developed a 
probability density for the mean microlens mass $\avgmstar$ (see the right panel of 
Figure~\ref{fig:velmass}) by convolving the prior on $v_{\rm e}$ with the probability 
density on $\hat{v}_{\rm e}$ from the simulation since $\hat{v}_{\rm e}=v_{\rm 
e}\avgmhat^{-1/2}$. To find the probability density for the source size in physical 
units, $d P(r_{\rm s}) / d \log{r_{\rm s}}$, we convolved the probability density for 
$\avgmstar$ with that of $\hat{r}_{\rm s}$. Assuming an inclination angle of 
$60^\circ$, we display the resulting distribution for the half-light radius 
($r_{1/2}$) in the right panel of Figure~\ref{fig:size}. In Fig.~\ref{fig:size}, we 
also show the probability density for $r_{1/2}$ using an assumed uniform prior on the 
mean microlens mass $0.1 \leq \avgmhat \leq 1.0$. While the two results are fully 
consistent, we promote the measurement without the uniform mass prior, the expectation 
value for which is $\log\{(r_{1/2}/{\rm cm})[\cos i/0.5]^{1/2}\} = 
15.4^{+0.3}_{-0.4}$. This measurement is the half-light radius of the quasar 
continuum source at the rest-frame effective wavelength $\lambda_{\rm{rest}}$ = 1930 
\AA\ (see the end of Sect.~\ref{sec:lcur}).

We also investigated the impact of uncertainty in the time delay on our measured 
microlensing sizes. We repeated our microlensing analysis at both the lower and upper 
1$\sigma$ bounds of the time delay, 41 and 52 d. We found size measurements of 
$\log\{(r_{1/2}/{\rm cm})[\cos i/0.5]^{1/2}\} = 15.5^{+0.3}_{-0.4}$ and 
$\log\{(r_{1/2}/{\rm cm})[\cos i/0.5]^{1/2}\} = 15.4^{+0.3}_{-0.3}$, respectively. 
These are fully consistent with the size measurement for the 47-d delay, demonstrating 
that time delay uncertainty has a small impact on the size. 

\section{Conclusions}
\label{sec:final}

We measured the virial mass of the SMBH at the centre of \object{SDSS J1339+1310} 
using optical-NIR spectra of image A. This resulted in ${\log \left( M_{\rm 
BH}/\rm{M_{\odot}} \right)}$ = 8.6 $\pm$ 0.4. Our black hole mass estimate is robust 
against the choice of the emission line and continuum luminosity, and it takes into
account the macrolens magnification, and the dust extinction in the Milky Way and 
lensing galaxy. Additional disregarded effects (e.g. extinction in the quasar host 
galaxy and microlens magnification) are expected to play a secondary role or even 
offset each other. While it is difficult to accurately quantify these effects, when 
correction factors ranging from 2/3 to 3/2 are applied to the denominator of 
Eq.~(\ref{eq1}), changes in the logarithm of the mass ($\la$ 0.1 dex) are well below 
its uncertainty. There are previous efforts to obtain virial black hole masses from 
samples of lensed quasars \citep[e.g.][]{Peng06,Gree10,Asse11}. The sixth and eighth 
columns of Table 5 in \citet{Asse11} show black hole mass estimates based on Balmer 
lines for a sample of twelve objects. A quarter of the 12 lensed quasars harbour SMBHs 
with a typical mass $\leq 4.5\times10^8 {\rm M_{\odot}}$, and \object{SDSS J1339+1310} 
belongs to this population of light objects. 

For a black hole of mass $M_{\rm BH}=4.0\times10^8 {\rm M_{\odot}}$, the gravitational 
radius is $r_{\rm g}=G M_{\rm BH} / c^2 = 5.9\times10^{13}$ cm. This black hole mass 
also determines the ISCO radius of the gas in a Schwarzschild geometry, which amounts 
to $r_{\rm ISCO} = 3.5\times10^{14}$ cm. In addition, the observed microlensing 
variability in the $r$ band allowed us to constrain the half-light radius of the 
region where the UV continuum at 1930 \AA\ is emitted. We found $\log\{(r_{1/2}/{\rm 
cm})[\cos i/0.5]^{1/2}\} = 15.4^{+0.3}_{-0.4}$, leading to a radial size of 
about $2.5\times10^{15}$ cm $\sim$ 7 $r_{\rm ISCO}$ for $i = 60^\circ$. 
Therefore, although UV observations are required to resolve the ISCO around the SMBH 
powering \object{SDSS J1339+1310}, our microlensing analysis resolves gas rings that 
are relatively close to the inner edge of the accretion disc. Moreover, using the 
simple thin-disc theory of \citet{Shak73}, our two measurements ($M_{\rm BH}$ and 
$r_{1/2}$) yielded an Eddington factor ${\log \left( L/\eta L_{\rm E} \right)}$ = 
0.8 $\pm$ 1.3 for $i = 60^\circ$. Despite the large uncertainty, the central value of 
this factor is consistent with a reasonable radiative efficiency $\eta \approx 
0.16\left( L/L_{\rm E} \right)$.
 
In Figure~\ref{fig:size}, we also display the probability density for the half-light 
radius of the accretion disc in \object{SDSS J1339+1310} scaled to $\lambda_{\rm 
rest}$ = 2500 \AA\ according to the standard thin-disc physics \citep{Shak73}. The 
expectation value for the 2500 \AA\ half-light radius $\log\{(r_{1/2}/{\rm cm})[\cos 
i/0.5]^{1/2}\} = 15.5^{+0.3}_{-0.4}$ is at 21$-$107 $r_{\rm g}$ for an inclination of 
$60^\circ$. This $1\sigma$ interval is barely consistent with the prediction of the 
accretion disc size - black hole mass relation \citep{Morg10} as updated by 
\citet{Morg18} 
\begin{equation}
\log[r_{1/2}/{\rm cm}]=(16.24\pm0.12) + (0.66\pm0.15)\log(M_{\rm BH}/10^9{\rm 
M_{\sun}}), 
\end{equation}
which predicts an accretion disc half-light radius at 2500 \AA\ of $106 \le 
r_{1/2}/r_{\rm g} \le 243$ for a black hole of mass $M_{\rm BH} = 4.0\times10^8 {\rm 
M_{\odot}}$. The minor discrepancy could easily be explained by the scatter in the 
\object{SDSS J1339+1310} black hole mass estimate. If the disc is more edge-on than 
our $60^\circ$ assumption, the inclination correction would also bring the measurement 
into closer agreement with the \citet{Morg18} scaling relation. However, it is 
important to realise that very high inclinations are unlikely in the presence of a 
dusty torus surrounding the gas disc \citep{Anto93}.

\begin{acknowledgements}
This paper is based on observations made with the Liverpool Telescope (LT) and the 
AZT-22 Telescope at the Maidanak Observatory. The LT is operated on the island of La 
Palma by Liverpool John Moores University in the Spanish Observatorio del Roque de los 
Muchachos (ORM) of the Instituto de Astrofisica de Canarias (IAC) with financial 
support from the UK Science and Technology Facilities Council. We thank the staff of 
the LT for a kind interaction before, during and after the observations. The Maidanak 
Observatory is a facility of the Ulugh Beg Astronomical Institute (UBAI) of the 
Uzbekistan Academy of Sciences, which is operated in the framework of scientific 
agreements between UBAI and Russian, Ukrainian, US, German, French, Italian, Japanese, 
Korean, Taiwan, Swiss and other countries astronomical institutions. We also present 
observations with the Nordic Optical Telescope (NOT) and the Italian TNG, operated on 
the island of La Palma by the NOT Scientific Association and the Fundaci\'on Galileo 
Galilei of the Istituto Nazionale di Astrofisica, respectively, in the Spanish ORM of 
the IAC. We also used data taken from several archives: National Optical Astronomy 
Observatory Science Archive, Pan-STARRS1 Data Archive, Sloan Digital Sky Survey Data
Releases, HST Data Archive, and ESO Science Archive Facility, and we are grateful to 
the many individuals and institutions who helped to create and maintain these public 
databases. This research has been supported by the MINECO/AEI/FEDER-UE grant 
AYA2017-89815-P and University of Cantabria funds to L.J.G. and V.N.S. This work was 
also supported by NSF award AST-1614018 to C.W.M. and M.A.C.
\end{acknowledgements}

\clearpage

\begin{appendix}

\section{Time delay through the updated dataset}
\label{sec:appena}

The light curves of the two images of \object{SDSS J1339+1310} display significant 
parallel variations on long timescales (see Fig.~\ref{fig:lcur}). Therefore, 
long-timescale variations must be closely related to quasar physics, providing an 
opportunity to estimate the time delay between A and B. We focused on the most densely 
sampled seasons, and as such, data in 2010$-$2011 were excluded for the time delay analysis. 
A reduced chi-square minimisation was then used to match the light curves of both 
images. In addition to a time delay $\Delta t_{\rm{AB}}$, we considered a magnitude 
offset $\Delta m_{\rm{AB}}$ for each of the nine seasons in 2009 and 2012$-$2019. 
These seasonal offsets account for long-timescale extrinsic (microlensing) 
variability. Our technique is fully described in Sect. 5.1 of \citetalias{Goic16}, and 
it relies on a comparison between the curve A and the time-shifted and binned curve B.
The time-shifted magnitudes of B are binned around the dates of A, and the semisize 
of bins is denoted by $\alpha$.  

\begin{figure}
\centering
\includegraphics[width=9cm]{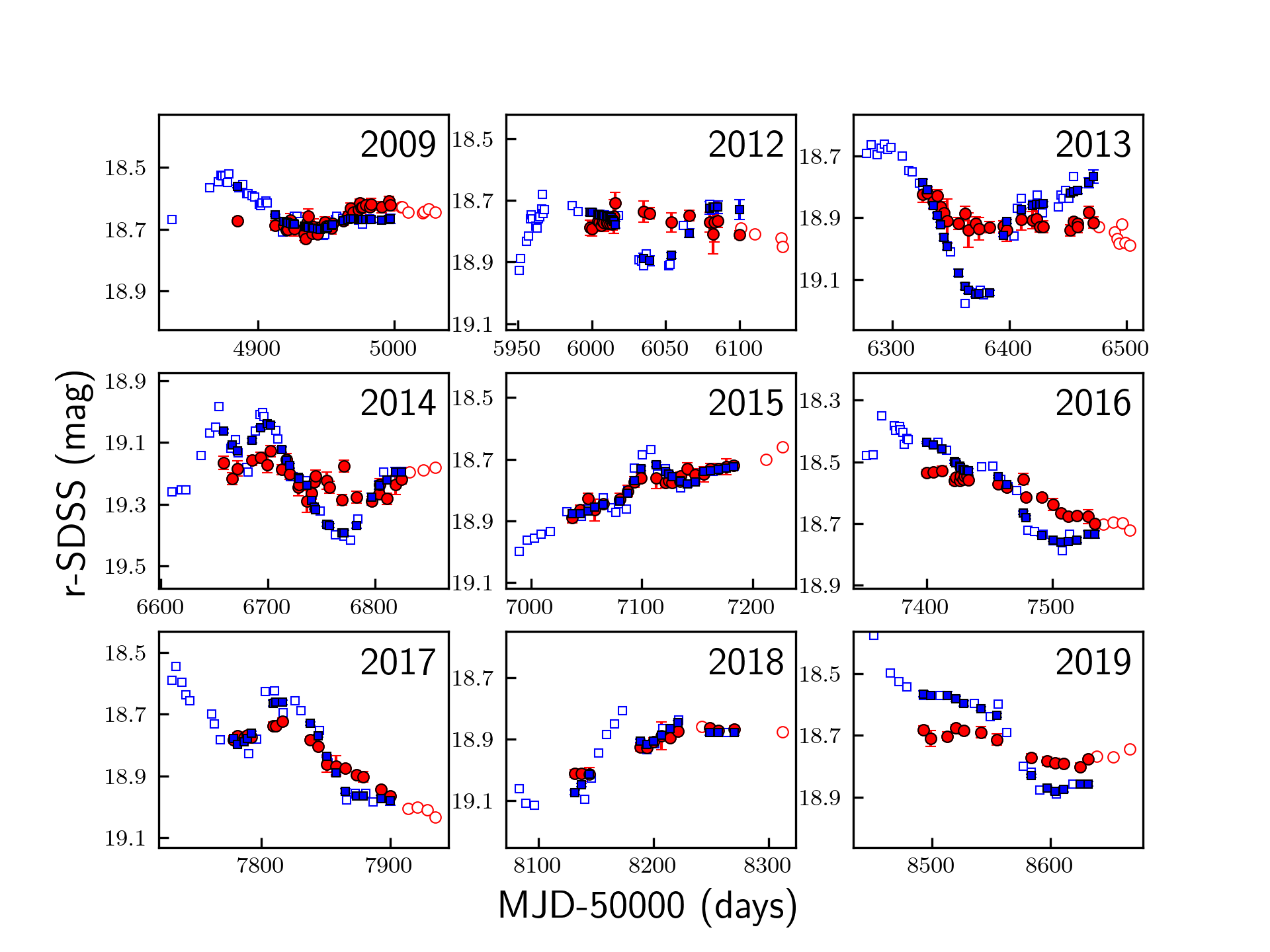}
\caption{Combined light curve in the $r$ band from the best solution in the third 
column of Table~\ref{tab:chi2bsol}. The A curve (red circles) is compared to the 
magnitude- and time-shifted B curve (blue squares). Open circles and squares are 
associated with the full brightness records, while filled circles and squares 
correspond to periods of overlap between both records. In the overlap periods, 
magnitudes of B are binned around dates of A using $\alpha$ = 20 d (filled squares; 
see main text).}
\label{fig:clcbsol}
\end{figure}

For $\alpha$ = 20 d, in Table~\ref{tab:chi2bsol}, we compare the best solution in 
\citetalias{Goic16} and the new best solution through the updated light curves. The 
combined light curve for this new solution is also shown in the nine panels of 
Fig.~\ref{fig:clcbsol}. As expected, there is a good agreement between the global 
trends of both images. Although a reasonably good agreement is also seen within some 
particular season (e.g. 2009, 2015 and 2018), short-timescale microlensing variability 
is present (e.g. intra-seasonal events/gradients in 2013$-$2014, 2016 and 2019). 
This rapid extrinsic variability is not taken into account in our reduced chi-square 
minimisation (see discussion below). From standard bins with $\alpha$ = 20 d, we also
found a 47-d solution that is practically as good as the one in the third column of 
Table~\ref{tab:chi2bsol}. Additionally, using 1000 pairs of synthetic curves 
\citepalias[see][]{Goic16}, we derived best solutions for $\alpha$ = 15, 20, and 25 d.
The distribution of 3000 delays resulting from simulated light curves and a reasonable 
range of $\alpha$ values, permited us to estimate uncertainties. Only delays within 
the interval from 46 to 50 d have an individual probability greater than 10\%, leading 
to an overall probability of 85\%, and hence a conservative 1$\sigma$ measurement  
$\Delta t_{\rm{AB}}$ = 48 $\pm$ 2 d. This delay interval basically coincides with the
1$\sigma$ measurement from the seasonal microlensing model in \citetalias{Goic16} 
($\Delta t_{\rm{AB}}$ = 47$^{+2}_{-1}$ d).  

In \citetalias{Goic16}, we also tried to account for all extrinsic variability using 
cubic splines and the PyCS software \citep{Tewe13,Bonv16}. However, such a method did 
not produce very robust results because it was not originally designed to model 
extrinsic variations that are as fast as or faster than intrinsic ones \citep{Tewe13}. 
Despite this fact, we obtained a 1$\sigma$ confidence interval $\Delta t_{\rm{AB}}$ = 
47$^{+5}_{-6}$ d using simulated light curves and a weak prior on the true time 
delay (i.e. it cannot be shorter than 30 d or longer than 60 d). Hence, the 
spline-like microlensing model led to a significantly enlarged error bar, which is 
adopted in this paper. We think this 1$\sigma$ interval is exaggeratedly large, and 
thus very conservative. First, considering only long-timescale extrinsic variability, 
a decade of monitoring observations with the LT and other telescopes strongly supports 
an error bar covering the delay range of 46$-$50 d (see above). This model exclusively 
has one non-linear parameter (time delay) and is appropriate. It accounts for 
extrinsic variations that distort the apparent long-term intrinsic signal, while the 
short-term extrinsic signal is assumed to be an extra-noise. Second, when all 
extrinsic variability is modelled, one deals with a complex non-linear optimisation, 
which may yield local minima, degeneracies, and so on. We have not yet found a 
fair way to simultaneously fit the delay and all microlensing activity. Despite 
these drawbacks, we are working on PyCS-based codes to account for the full 
microlensing signal in \object{SDSS J1339+1310}-like systems. These codes could 
produce robust results in a near future.

\begin{table}
\centering
\caption{Best solutions for $\alpha$ = 20 d.}
\begin{tabular}{lcc}
\hline\hline
 & \citetalias{Goic16} & This paper \\
\hline
$\Delta t_{\rm{AB}}$ (d)    &    47 &    48 \\
$\Delta m_{\rm{AB}}(2009)$  & 0.349 & 0.348 \\
$\Delta m_{\rm{AB}}(2012)$  & 0.419 & 0.417 \\
$\Delta m_{\rm{AB}}(2013)$  & 0.133 & 0.139 \\
$\Delta m_{\rm{AB}}(2014)$  & 0.118 & 0.116 \\
$\Delta m_{\rm{AB}}(2015)$  & 0.304 & 0.302 \\
$\Delta m_{\rm{AB}}(2016)$  &  ---  & 0.189 \\
$\Delta m_{\rm{AB}}(2017)$  &  ---  & 0.159 \\
$\Delta m_{\rm{AB}}(2018)$  &  ---  & 0.261 \\
$\Delta m_{\rm{AB}}(2019)$  &  ---  & 0.136 \\
\hline
\end{tabular}
\tablefoot{
The time-dependent magnitude offset $\Delta m_{\rm{AB}}(t) = m_{\rm{A}}(t) - 
m_{\rm{B}}(t + \Delta t_{\rm{AB}})$ is assumed to be constant within each season. All
seasonal offsets are positive because we use the opposite sign to that in 
\citetalias{Goic16}.    
} 
\label{tab:chi2bsol}
\end{table} 

\end{appendix}

\end{document}